\documentclass[journal=mamobx,manuscript=article]{achemso}
\setkeys{acs}{articletitle = true}

\usepackage{chemformula,amsmath,amssymb} 
\usepackage[T1]{fontenc} 
\usepackage{adjustbox}
\usepackage{changes,cancel}
\definechangesauthor[color=red]{Rev1}
\definechangesauthor[color=green]{Rev2}
\definechangesauthor[color=blue]{Us}
\author{Mattia Alberto Ubertini}
\affiliation{Scuola Internazionale Superiore di Studi Avanzati (SISSA), Via Bonomea 265, 34136 Trieste, Italy}
\email{mubertin@sissa.it}
\author{Jan Smrek}
\affiliation{Faculty of Physics, University of Vienna, Boltzmanngasse 5, A-1090 Vienna, Austria}
\email{jan.smrek@univie.ac.at}
\author{Angelo Rosa}
\affiliation{Scuola Internazionale Superiore di Studi Avanzati (SISSA), Via Bonomea 265, 34136 Trieste, Italy}
\email{anrosa@sissa.it}

\title{Entanglement length scale separates threading from branching of unknotted and non-concatenated ring polymers in melts}

\abbreviations{IR,NMR,UV}

\begin{document}








\begin{abstract}
Current theories on the conformation and dynamics of unknotted and non-concatenated ring polymers in melt conditions describe each ring as a tree-like double-folded object. While evidence from simulations supports this picture on a single ring level, other works show pairs of rings also thread each other -- a feature overlooked in the tree theories. Here we reconcile this dichotomy using Monte-Carlo simulations of the ring melts with different bending rigidity. We find that rings are double-folded (more strongly for stiffer rings) {\it on and above} the entanglement length scale, while the threadings are localized {\it on smaller scales}. The different theories disagree on the details of the tree structure, {\it i.e} the fractal dimension of the backbone of the tree. In the stiffer melts we find an indication of a self-avoiding scaling of the backbone, while more flexible chains do not exhibit such a regime. 
Moreover, the theories commonly neglect threadings, and assign different importance to the impact of the progressive constraint release (tube dilation) on single ring relaxation due to the motion of other rings. Despite each threading creates only a small opening in the double-folded structure, the threading loops can be numerous and their length can exceed substantially the entanglement scale. We link the threading constraints to the divergence of the relaxation time of a ring, if the tube dilation is hindered by pinning a fraction of other rings in space. Current theories do not predict such divergence and predict faster than measured diffusion of rings, pointing at the relevance of the threading constraints in unpinned systems as well. Revision of the theories with explicit threading constraints might elucidate the validity of the conjectured existence of topological glass.
\end{abstract}


\section{1. Introduction\label{sec:Intro}}
Topological constraints emerging from the mutual uncrossability between distinct chain segments dominate the viscoelastic behavior of polymer systems in high-density (melt) conditions~\cite{DeGennes1979,DoiEdwardsBook,RubinsteinColbyBook}.
In this context, of notable interest are those situations where polymers are prepared in a well-defined topological state which remains quenched as polymers diffuse and flow.
The simplest example in this respect, and the central topic of the present work, is the case of melts of {\it unknotted} and {\it non-concatenated}  
ring polymers.
People have been now studying this particular class of polymer solutions for several decades, from the
theoretical~\cite{KhokhlovNechaev1985,Rubinstein1986,CatesDeutschJPhysFrance1986,Rubinstein1994,Cates2000,halverson2011molecular-statics,halverson2011molecular-dynamics,SakauePRL2011,HalversonPRL2012,GrosbergSM2014,RosaEveraersPRL2014,ObukhovWittmer2014,SmrekGrosbergRingsR015,SmrekGrosbergACSMacroLett2016,RosaEveraersJCP2016,PanyukovRubinsteinMacromolecules2016,MichielettoSM2016,DellSchweizer2018,SmrekRosa2019,SchramRosaEveraers2019,rosa2019conformational,MichielettoSakaue2021,OConnorPRL2020,ElhamPRE2021}
as well as the
experimental~\cite{Kapnistos2008,Pasquino2013,Iwamoto2018,VlassopoulosPRL2019,KrutevaACSML2020,KrutevaPRL2020}
point of view.
Researchers have shown that there exist intriguing conceptual connections between melts of ring polymers and, for instance, chromosomes~\cite{RosaEveraersPlos2008,HalversonSmrekRPP_2014} and a polymer glass based on the nonlinear topology of the chains~\cite{Turner2013,Michieletto2016,MichielettoNahaliRosa2017,SmrekChubak2020}.
Yet, because of the particular nature of the problem,
many fundamental aspects concerning the physics of ring polymer melts remain poorly understood.
At present in fact, physical theories taking exactly into account the constraint by means of suitably topological invariants~\cite{BreretonVilgis1995,MorozKamien1997,KholodenkoVilgis1998,KungKamien2003,Ferrari2019} remain mathematical hard, if not completely intractable, problems
and their applicability to the dense, many-chain systems is limited.
Therefore, the various physical pictures which have been proposed so far~\cite{KhokhlovNechaev1985,Rubinstein1986,Rubinstein1994,ObukhovWittmer2014,PanyukovRubinsteinMacromolecules2016} introduce suitable approximations to deal with the constraint which make the problem more affordable but, inevitably, they require a supplement of validation, either from experiments or from numerical simulations.

In a series of landmark papers~\cite{KhokhlovNechaev1985,Rubinstein1986,Rubinstein1994} topological constraints have been approximated, in a mean-field fashion, as a lattice of infinitely thin impenetrable obstacles (see Fig.~\ref{fig:DoubleFoldedRings}), therefore, in avoiding concatenation, rings should protrude through them by {\it double-folding} on {\it branched, tree-like} conformations. 
As a consequence~\cite{GrosbergSM2014,RosaEveraersJCP2016,rosa2019conformational}, rings form compact shapes whose mean linear size or gyration radius, $\langle R_g \rangle$, scales with the total contour length of the chain, $L$, as: 
\begin{equation}\label{eq:IntroducingRg}
\langle R_g \rangle \sim d_T \! \left( \frac{L}{L_e} \right)^{1/3} \, ,
\end{equation}
for $L$ larger than the characteristic and material-dependent~\cite{EveraersCommodity2020} contour length scale, $L_e$, known as the {\it entanglement length} and with $d_T$, the {\it tube diameter} of the melt, equivalent to the average mesh distance of the array of topological obstacles~\cite{DeGennes1979,DoiEdwardsBook,RubinsteinColbyBook}.
Many non-trivial predictions of the lattice-tree model -- especially single-chain ones as, for instance, the scaling behavior of the rings (Eq.~\eqref{eq:IntroducingRg}) or their monomer diffusion in the melt -- appear in good agreement with brute-force dynamic simulations~\cite{CatesDeutschJPhysFrance1986,Cates2000,halverson2011molecular-statics,halverson2011molecular-dynamics,SmrekGrosbergRingsR015,MichielettoSM2016,rosa2019conformational}.
Notably, the model has also helped casting a multi-scale algorithm~\cite{RosaEveraersPRL2014} for generating ring melt conformations at negligible computational cost.

While seemingly accurate, the double-folding/branching model (Fig.~\ref{fig:DoubleFoldedRings}) dismisses~\cite{SmrekRosa2019,SchramRosaEveraers2019} the possibility of ring-ring interpenetrations (or, {\it threadings}) which instead -- and without violating the non-concatenation constraint! -- have physical consequences which become particularly evident when melts are driven out-of-equilibrium, such as in an elongation flow~\cite{OConnorPRL2020} or for an induced asymmetry in the local monomer mobilities~\cite{SmrekChubak2020}.
Moreover, from a broader perspective it remains unclear if the entanglement scales $L_e$ and $d_T$ (Eq.~\eqref{eq:IntroducingRg}) are the only relevant ones for melts of rings or if, and up to what extent, they are influenced by the local bending rigidity or Kuhn length~\cite{RubinsteinColbyBook} $\ell_K$ of the chain. 
In fact, the value of $L_e$ is measured from a linear melt ({\it e.g.} by primitive path analysis~\cite{EveraersScience2004}) and the length scale is only {\it assumed} to be applicable for rings (with the same polymer model) as well, although no {\it direct} method to find the value explicitly from the rings is known.
Some recent comparison of crazing in linear and ring glasses suggests~\cite{Ge_crazing_2021,Smrek_commentary2022} that $L_e({\rm rings}) / L_e({\rm linear}) \approx 4$, but its role in equilibrium ring melts, as well as its connection to $\ell_K$, has not been thoroughly investigated.
All these aspects (double-folding/branching, threadings, entanglement scales) are clearly all related to each other and, yet, how they mutually influence each other remain poorly understood.

In order to address these questions, in this work we perform a systematic investigation of the static and dynamic properties of melts of unknotted and non-concatenated ring polymer conformations obtained in large-scale computer simulations.
To this purpose, we present first a modified version of the kinetic Monte Carlo algorithm described in~\cite{Hugouvieux2009,Schram-LatticeModel2018} with the purpose of achieving higher values of the polymer Kuhn length $\ell_K$: in doing so, we built on the numerically efficient strategy originally devised  in~\cite{Cates2000} which allows us to increase substantially the overlap between polymer chains, and to reach feasibly the asymptotic regime Eq.~\eqref{eq:IntroducingRg}, with relatively moderate chain sizes and, hence, computing times.
Then, by
employing established tools like the polymer
mean gyration radius and shape~\cite{BishopMichels1986},
contour length correlations~\cite{Cates2000} and spatial contacts~\cite{Dekker_HiC2009},
spanned surface~\cite{SchramRosaEveraers2019} and minimal surface~\cite{LangMacromolecules2013,SmrekGrosbergACSMacroLett2016},
alongside the motion of the melt for unperturbed conditions~\cite{KremerGrest-JCP1990} or after partial pinning~\cite{Michieletto2016} of a given fraction of the polymer population,
we find out that rings are systematically threading each others {\it via} the formation of locally double-folded structures on length scales below the tube diameter $d_T$.
Additionally, we show that the threading loops of contour length $<L_e$ (which we name {\it shallow threadings}) influence significantly many, non-universal, properties of the rings.

The paper is structured as follows.
In Section~2, 
we present the notation employed in this work (Sec.~2.1) 
and summarize the main predictions for the structure (Sec.~2.2) 
and the dynamics (Sec.~2.3) 
of melts of rings according to various models with the emphasis on their lattice-tree features.
In Sec.~3, 
we present and discuss our lattice polymer model and derive the relevant length and time scales of the polymer melts.
In Sec.~4 
we present the main results, for the structure (Sec.~4.1) 
and the dynamics (Sec.~4.2) 
of the rings.
In these sections and in the final Sec.~5 
we present and discuss how the tree-like models can be reconciled with the threading features. We also indicate measurements that can help to discriminate between the different models and discuss the common deficiency of all of them.
Additional tables and figures are included in the Supporting Information (SI).

\section{2. Ring polymers in melt: theoretical background\label{sec:Theory}}

\subsection{2.1 Entanglement length and time units: definitions and notation}\label{sec:Theory-LengthTimeScales}
Single polymers in melt are made of a linear sequence of monomer units with mean bond length $=\langle b \rangle$.
The total number of monomers of each polymer chain is $=N$ and the polymer contour length is $L=N\langle b\rangle$, while we denote by $\ell \, (\leq L)$ the contour length of a polymer sub-chain made of $n \equiv \ell / \langle b\rangle \leq N$ monomers.

Topological constraints (entanglements) affect polymer conformations in melt when the total contour length of each chain, $L$, exceeds the characteristic entanglement length scale $L_e$.
In general $L_e$ is a non-trivial function of the chain bending stiffness or the Kuhn length, $\ell_K$, and the Kuhn segment density, $\rho_K$, of the melt (see Eq.~\eqref{eq:Uchida-etal} and Ref.~\cite{EveraersScience2004}).
Then, the mean gyration radius of the polymer chain of contour length $L=L_e$ and Kuhn length $\ell_K$,
\begin{equation}\label{eq:DefineTubeDiameter}
d_T \equiv \sqrt{ \frac{\ell_K L_e}6 } \, ,
\end{equation}
is of the order of the cross-sectional diameter of the {\it tube-like} region~\cite{DeGennes1979,DoiEdwardsBook,RubinsteinColbyBook} where the polymer is confined due to the presence of topological constraints.
Unless otherwise said, we will express chain contour lengths $L$ in units of the entanglement length $L_e$ and, to this purpose, we introduce the compact notations~\cite{RosaEveraersPRL2014,SchramRosaEveraers2019}:
\begin{eqnarray}
Z \equiv \frac{L}{L_e} & \equiv & \frac{N}{N_e} \, , \label{eq:LoverLe=Z} \\
z \equiv \frac{\ell}{L_e} & \equiv & \frac{n}{N_e} \, , \label{eq:elloverLe=z}
\end{eqnarray}
where $N_e$ corresponds to the number of monomers $L_e / \langle b \rangle$ of an entanglement length and
$Z$ and $z$ are the number of entanglements of the polymer chain and the polymer sub-chain of contour lengths $L$ and $\ell$, respectively~\cite{MotivationForLk}. 
Accordingly, variables such as the polymer mean-gyration radius $\langle R_g \rangle$ or the mean-magnetic radius $\langle R_m \rangle$ (see definitions Eqs.~\eqref{eq:Rg2} and~\eqref{eq:MagnRadius}, respectively) are expressed in units of the tube diameter $d_T$ (Eq.~\eqref{eq:DefineTubeDiameter}).

With respect to polymer dynamics, entanglements affect the motion of the chains on time scales $\tau$ larger than a characteristic time $\tau_e$, known as the {\it entanglement time}.
$\tau_e$ is defined as in~\cite{KremerGrest-JCP1990} namely it corresponds to the time scale where the monomer mean-square displacement $g_1(\tau=\tau_e)$ (Eq.~\eqref{eq:Introduce-g1}) is $\simeq 2d_T^2$ (Eq.~\eqref{eq:DefineTubeDiameter}).
From now on, 
we adopt $\tau_e$ as our main unit of time.

In Sec.~3.4, 
we present a detailed derivation of these quantities for the polymer melts considered in this work.

\begin{figure}
\includegraphics[width=0.45\textwidth]{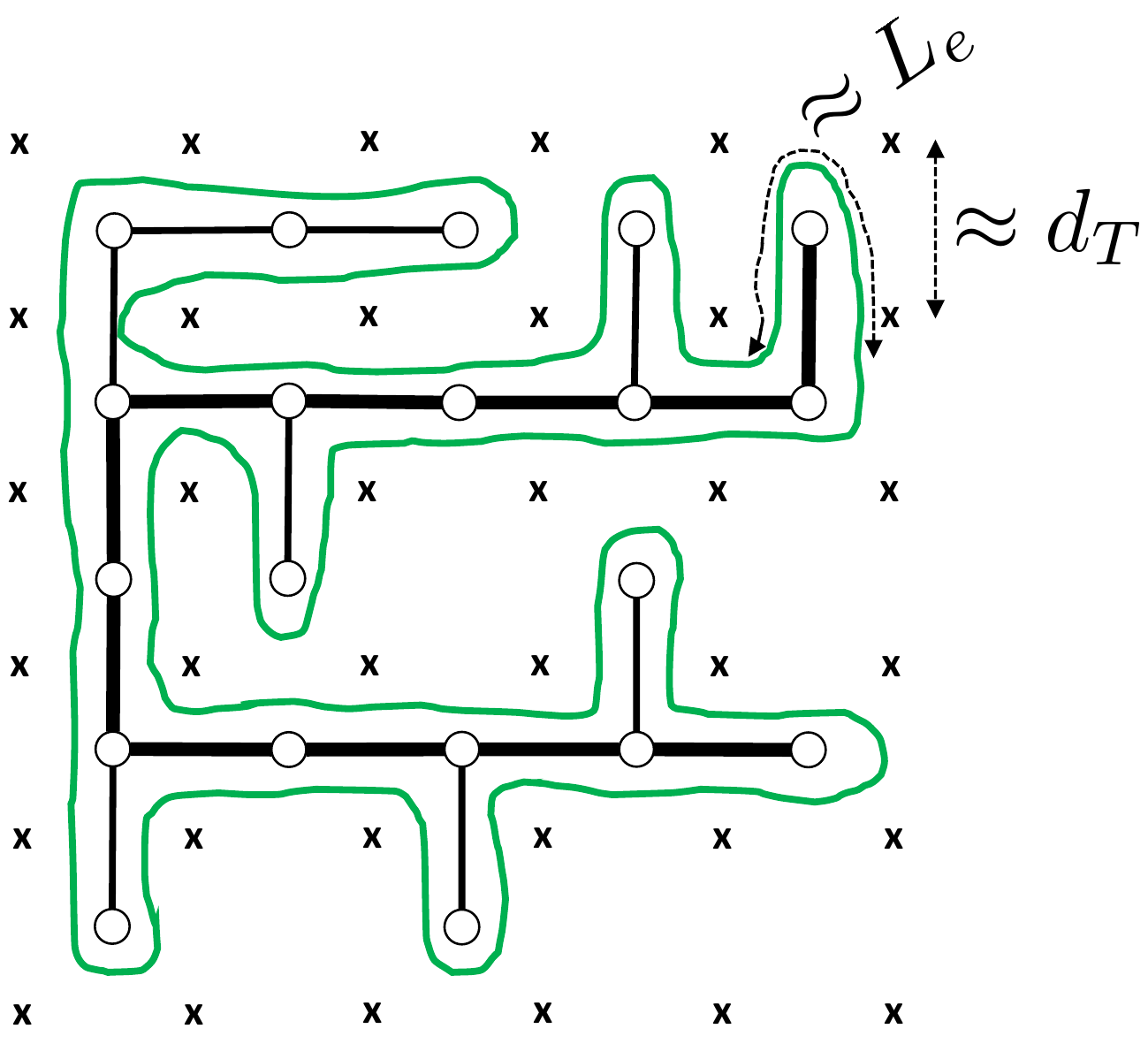}
\caption{\label{fig:DoubleFoldedRings}
Schematic picture of a single ring polymer conformation (green line) in melt which double-folds around its tree-like backbone.
The crosses $(\times)$ represent the lattice of topological obstacles created by the surrounding rings in the melt.
The lattice spacing is of the order of the tube diameter $d_T$, and the polymer contour length inside each unit cell is $\approx L_e$.
Random displacements of the loops protruding from the mean path on the backbone of the branched structure (thick black line) changes ring's shape and dominates melt dynamics at long time scales~\cite{Rubinstein1994,SmrekGrosbergRingsR015}.
}
\end{figure}
%

\subsection{2.2 Ring structure\label{sec:Theory-Structure}}
According to the lattice-tree picture~\cite{KhokhlovNechaev1985,Rubinstein1986,Rubinstein1994,RosaEveraersPRL2014,GrosbergSM2014,SmrekGrosbergRingsR015}, ring conformations in melt are the result of the balance between compression to avoid linking with other rings and swelling to avoid self-knotting.
At the same time, this size competition can be viewed as a balance between double-folding (which minimizes threadings between chains) and random branching (which maximizes polymer entropy).
As a consequence, a melt of rings can be mapped~\cite{RosaEveraersPRL2014} to an equivalent melt of randomly branching polymers or lattice trees with the same large-scale behavior, see Fig.~\ref{fig:DoubleFoldedRings}.
In turn, this mapping can be employed to derive quantitative predictions~\cite{GrosbergSM2014,RosaEveraersJCP2016,EveraersFloryBP2017,rosa2019conformational} for the scaling exponents describing the characteristic power-law behaviors as a function of $L/L_e \equiv Z \gtrsim 1$ of the following observables:
\begin{itemize}
\item[(i)] The mean ring size or gyration radius (Eq.~\eqref{eq:IntroducingRg}) as a function of the ring mass:
\begin{equation}\label{eq:Rings-RvsZ}
\langle R_g \rangle \sim d_T \, Z^\nu \, , \, \, \, \nu=1/3 \, ;
\end{equation}
\item[(ii)] The mean path length on the backbone of the tree (Fig.~\ref{fig:DoubleFoldedRings}, thick black line) as a function of the ring mass:
\begin{equation}\label{eq:Rings-LvsZ}
\langle L_{\rm tree} \rangle \sim d_T \, Z^\rho \, , \, \, \, \rho=5/9 \, ;
\end{equation}
\item[(iii)] The mean ring size as a function of the mean path length:
\begin{equation}\label{eq:Rings-RvsL}
\langle R_g \rangle \sim d_T^{1-\nu_{\rm path}} \langle L_{\rm tree} \rangle^{\nu_{\rm path}} \, , \, \, \, \nu_{\rm path}=3/5 \, .
\end{equation}
\end{itemize}
Eq.~\eqref{eq:Rings-RvsZ}  means that rings ({\it i.e.} the equivalent trees) behave like compact, space-filling objects while Eq.~\eqref{eq:Rings-RvsL} expresses the fact that linear paths follow self-avoiding walk statistics~\cite{GrosbergSM2014}.
Notice also that Eq.~\eqref{eq:Rings-RvsL} follows directly from Eqs.~\eqref{eq:Rings-RvsZ} and~\eqref{eq:Rings-LvsZ}, hence $\nu_{\rm path} = \nu / \rho$.
While the model explains and quantifies the origin of branching and correctly predicts the measured exponent $\nu$ in the asymptotic limit, there have been no attempts so far to measure the exponents $\rho$ or $\nu_{\rm path}$ in simulations directly (the backbone is not easily ``visible'' in the conformations).
Moreover, the theory is based on scales above $N_e$ and states explicitly~\cite{SmrekGrosbergRingsR015} that the branches might not be double-folded to the monomer scale. This means that the branches can open up to sizes about $d_{T}$ to allow for a branch of another ring to thread through. As we will show in the Sec.~4.1 
this is indeed the case.

Another very popular and accurate model is the so called fractal loopy globule (FLG)~\cite{PanyukovRubinsteinMacromolecules2016}.
The model is based on the conjecture that the ring structure arises from a constant (and equal to the Kavassalis-Noolandi number~\cite{KavassalisNoolandiPRL1987}) overlap parameter (number of segments/loops of a given length sharing a common volume) on all scales above $L_e$ in a self-similar manner.
As a result, the exponent $\nu=1/3$ is practically postulated. As such, the ring structure, as loops on loops, is ``somewhat analogous" to a randomly branched structure, where the loops of the FLG are viewed as the branches.
The authors assert that the loops are not perfect double-folded, but do not elaborate on the double-folded structure further, except that the number of loops/branches $n(r)$ of size $r$ per ring is $n(r) \sim r^{-3}$.

The difference of the FLG and the lattice-tree models, from the structural point of view is in the branching statistics, manifested in the structure of the mean (primitive) path\footnote{Note that here we use the terms primitive path and tree backbone interchangeably, because they have the same meaning for linear polymers -- the shortest end-to-end path of the chain to which its contour can be contracted without crossing other chains. In rings, there are no ends and therefore such equivalence is, to say the least, not clear. The tree backbone, governing the stretching and branching of the ring is a properly weighted average of all possible path lengths in the tree structure (see~\cite{SmrekGrosbergRingsR015} for details), while the primitive path of the ring is measured in~\cite{PanyukovRubinsteinMacromolecules2016} with a method analogous, but not identical, to primitive path analysis~\cite{EveraersScience2004}.}.
While in the lattice-tree model the scaling of the size of a segment of the ring with the corresponding primitive path is governed by the exponent $\nu_{\rm path}$ on scales above $N_e$~\eqref{eq:Rings-RvsL}, in the FLG, the scaling is more complicated because it takes into account ``tube dilation''.
The tube dilation means freeing of the constraints imposed by the other surrounding segments due to their motion, hence effectively increasing the length scale between constraints, making it time-dependent $N_e(t)$.
In the FLG model if the segment is shorter than $N_e(t)$ its primitive path is just straight, hence the size of the segment scales linearly with the length.
If the segment is longer than $N_e(t)$, its size scales with the exponent $\nu=1/3$ equal to that of the whole ring.
The analysis of the primitive paths in~\cite{PanyukovRubinsteinMacromolecules2016} reports that they do not observe the scaling of the size of the primitive path with its length with the exponent $\nu_{\rm path}=3/5$.
Yet their analysis focuses on rather flexible system only and the reported dependence broadly and smoothly crosses over from the exponent $1$ to $1/3$ (hence visiting all the intermediate exponents). As we show below in Sec.~4.1 
we find an indication of the exponent $\nu_{\rm path}=3/5$ in the stiff melts.


\subsection{2.3 Ring dynamics\label{sec:Theory-Dynamics}}
As originally pointed out by Obukhov {\it et al.}~\cite{Rubinstein1994}, the branched structure induced by the topological obstacles (Fig.~\ref{fig:DoubleFoldedRings}) has direct implications for ring dynamics as well.
Here, we limit ourselves to a schematic description of the physics of the process and to recapitulate the main results, without entering into a detailed derivation which, being far from trivial, the interested reader can find explained in detail in~\cite{Rubinstein1994,SmrekGrosbergRingsR015,ElhamPRE2021}.

In order to quantify chain dynamics we introduce~\cite{KremerGrest-JCP1990} the monomer mean-square displacement
\begin{equation}\label{eq:Introduce-g1}
g_1(\tau) \equiv \left\langle \frac1N \sum_{i=1}^N ( \vec r_i(\tau+t) - \vec r_i(t) )^2 \right\rangle \, ,
\end{equation}
and
the mean-square displacement of the ring centre of mass
\begin{equation}\label{eq:Introduce-g3}
g_3(\tau) \equiv \left\langle ( \vec r_{\rm cm}(\tau+t) - \vec r_{\rm cm}(t) )^2 \right\rangle \, .
\end{equation}
as a function of time $\tau$.
Then, three regimes can be identified.
For time scales $\tau \lesssim \tau_e$ and length scales $\lesssim d_T$, monomer motion is not affected by entanglements and we expect the characteristic Rouse-like~\cite{RubinsteinColbyBook} behavior,
$g_1(\tau) \sim \tau^{1/2}$
and
$g_3(\tau) \sim \tau$.
On intermediate time scales $\tau \gtrsim \tau_e$, the motion of the ring is dominated by mass transport along the mean path on the tree-like backbone (Fig.~\ref{fig:DoubleFoldedRings}).
Finally, for large time scales $\tau \gtrsim \tau_r$ where $\tau_r \approx \tau_e \, Z^{2+\rho}$ is the global relaxation time of the chain~\cite{SmrekGrosbergRingsR015}, the whole chain is simply diffusing, {\it i.e.} $g_1(\tau) \sim g_3(\tau) \sim \tau$.
The complete expressions for $g_1$ and $g_3$, up to numerical prefactors but including the coefficients for smooth crossovers~\cite{SmrekGrosbergRingsR015}, are given by:
\begin{equation}\label{eq:LatticeTree-g1-Complete}
g_1 \sim d_T^2 \times \left\{
\begin{array}{lc}
\left( \frac{\tau}{\tau_e} \right)^{1/2} , & \tau \lesssim \tau_e \\
\\
\left( \frac{\tau}{\tau_e} \right)^{2\nu / (\rho+2)} , & \tau_e \lesssim \tau \lesssim \tau_r \\
\\
Z^{2\nu -\rho-2} \, \frac{\tau}{\tau_e}, & \tau \gtrsim \tau_r
\end{array}
\right.
\end{equation}
and
\begin{equation}\label{eq:LatticeTree-g3-Complete}
g_3 \sim d_T^2 \times \left\{
\begin{array}{lc}
\frac1Z \frac{\tau}{\tau_e} , & \tau \lesssim \tau_e \\
\\
\frac1Z \left( \frac{\tau}{\tau_e} \right)^{(2\nu+1) / (\rho+2)} , & \tau_e \lesssim \tau \lesssim \tau_r \, , \\
\\
Z^{2\nu -\rho-2} \, \frac{\tau}{\tau_e}, & \tau \gtrsim \tau_r
\end{array}
\right.
\end{equation}
where the exponents $\nu$ and $\rho$ appearing here are the same ones introduced in the scaling behaviors of the static quantities, see Sec.~2.2.

Finally, we introduce the monomer mean-square displacement in the frame of the chain center of mass:
\begin{equation}\label{eq:Introduce-g2}
g_2(\tau) \equiv \left\langle \frac1N \sum_{i=1}^N ( \vec r_i(\tau+t) - \vec r_{\rm cm}(t+\tau) - \vec r_i(t) + \vec r_{\rm cm}(t) )^2 \right\rangle \, .
\end{equation}
It is not difficult to see that $g_2(\tau) \simeq g_1(\tau) - g_3(\tau)$ and that $g_2(\tau\rightarrow\infty) = 2\langle R_g^2\rangle$ where $\langle R_g^2\rangle$ is the chain mean-square gyration radius (see definition, Eq.~\eqref{eq:Rg2}).
We adopt the appearance of a plateau in the large-time behavior of $g_2$ as the signature that our chains have attained complete structural relaxation (see Fig.~S1 in SI). 

The FLG model~\cite{PanyukovRubinsteinMacromolecules2016} takes into account the tube dilation, as mentioned in the Sec.~2.1, 
but otherwise uses self-similar dynamics characteristic for other branched ring conformations as well.
Therefore, as shown in\cite{PanyukovRubinsteinMacromolecules2016}, the predictions for the $g_{1}$ of both models, the lattice-tree dynamics and the FLG, can be concisely written as
\begin{equation}\label{eq:g1_FLG_and_LT}
g_{1}(\tau) \sim \tau^{2 / ((2/\nu) +(1-\theta)/\nu_{\rm path}+\theta)} \, ,
\end{equation}
for times $\tau_{e} \leq \tau \leq \tau_r$. The parameter $\theta$ governs the tube dilation: $\theta = 0$ means no tube dilation as in the lattice-tree model, while $\theta=1$ means full tube dilation as in the FLG model.
Note that in FLG, because of the full tube dilation the exponent $\nu_{\rm path}$ does not impact the dynamics. The exponents are $0.26$ for the lattice-tree model and about $0.29$ for FLG.

Interestingly both, the FLG and the annealed lattice-tree model, are very close in the predictions of the scaling of the monomer mean-squared displacements and other dynamical quantities~\cite{PanyukovRubinsteinMacromolecules2016}, apart from the scaling of the diffusion coefficient with $N$.
This is computed as $D \sim R^2 / \tau_r$,
where the relaxation time $\tau_{r}$ involving the tube dilation can be written as
\begin{equation}\label{eq:tau_relax-FLG}
\tau_r \sim \tau_e (N/N_e)^{2+ (1-\theta)\nu/\nu_{\rm path} + \theta\nu} \, ,
\end{equation}
which gives
\begin{equation}\label{eq:D_FLG_and_LT}
D \sim N^{-2 + \nu(2-\theta-(1-\theta)/\nu_{\rm path})} \, .
\end{equation}
Although the exponent is difficult to measure accurately in simulations due to a need of very long runs, the theories underestimate the exponent obtained from the simulations~\cite{halverson2011molecular-dynamics,PanyukovRubinsteinMacromolecules2016} as well as that from the experiments~\cite{KrutevaPRL2020} by $0.67$ and $0.45$ for the FLG and the lattice-tree model, respectively.
As we show below in Sec.~4.2 
the source of the discrepancy can be caused by the threadings.


\section{3. Polymer model and numerical methods\label{sec:ModelMethods}}

\subsection{3.1 Melts of rings: the kinetic Monte Carlo algorithm\label{sec:PolymerModel}}
We model classical solutions of ring polymers of variable Kuhn length $\ell_K$ and with excluded volume interactions by adapting the kinetic Monte Carlo (kMC) algorithm for elastic lattice polymers on the three-dimensional face-centered-cubic (fcc) lattice introduced originally in~\cite{Hugouvieux2009,Schram-LatticeModel2018} and employed later in several studies on polymer melts~\cite{OlartePlata2016,SchramRosaEveraers2019,ubertini2021computer}.
In the following we provide the main features of the model, while we refer the reader to the mentioned literature for more details.

In the model, any two consecutive monomers along the chain sit either on nearest-neighbor lattice sites or on the same lattice site (with no more than two consecutive monomers occupying the same lattice site), while nonconsecutive monomers are never allowed to occupy the same lattice site due to excluded volume.
By adopting the lattice distance $a$ between fcc nearest-neighbor sites as our unit distance, the bond length $b$ between nearest-neighbor monomers fluctuates between $a$ and $0$ (the latter case corresponding to a unit of {\it stored length}):
for an average bond length $=\langle b\rangle$, a polymer chain with $N$ bonds has then a total contour length $L = N\langle b\rangle < Na$. 
Thanks to this numerical ``trick'', polymers are made effectively elastic~\cite{Hugouvieux2009}. 

Ring polymers in melt are asymptotically compact (Eq.~\eqref{eq:IntroducingRg}), yet reaching this regime requires the simulation of very large rings which, in turn, may imply prohibitively long~\cite{Cates2000,RosaEveraersPRL2014,halverson2011molecular-statics} equilibration times.
In order to overcome this limitation and, yet, still achieving substantial overlap between the different polymer chains for moderate chain lengths and, hence, feasible simulation times,
we adopt the efficient strategy described in~\cite{Cates2000} and consider polymer chains which are locally stiff, namely polymers whose Kuhn length~\cite{RubinsteinColbyBook} $\ell_K$ 
is significantly larger than the mean bond length $\langle b\rangle$.
To this purpose, we have complemented the chain Hamiltonian by introducing the bending energy term:
\begin{equation}\label{eq:CosPotential}
\frac{\mathcal H_{\rm bend}}{k_BT} = -\kappa_{\rm bend} \sum_{i=1}^{L / a} \cos \theta_i \equiv -\kappa_{\rm bend} \sum_{i=1}^{L / a} \frac{\vec{t}_i \cdot \vec{t}_{i+1}}{|\vec{t}_i| |\vec{t}_{i+1}|} \, ,
\end{equation}
where
$\kappa_{\rm bend}$ represents the bending stiffness which determines $\ell_K$ (see Sec.~3.4) 
and
$\vec t_i \equiv \vec r_{i+1}-\vec r_{i}$ is the oriented bond vector between monomers\footnote{For ring polymers, it is implicitly assumed the periodic boundary condition along the chain $N+1\equiv 1$.} 
$i$ and $i+1$ having spatial coordinates $\vec r_i$ and $\vec r_{i+1}$.
Importantly, since bond vectors are obviously ill-defined when two monomers form a stored length, the sum in Eq.~\eqref{eq:CosPotential} is restricted to the {\it effective} bonds of the chains.
By increasing $\kappa_{\rm bend}$, the energy term Eq.~\eqref{eq:CosPotential} makes polymers stiffer. 

Then, the dynamic evolution of the melts proceeds according to the following Metropolis-Hastings-like~\cite{Metropolis1953} criterion.
One monomer is picked at random and displaced towards one of the nearest lattice sites.
The move is accepted based on the energy term Eq.~\eqref{eq:CosPotential} and if, at the same time, {\it both} chain connectivity and excluded volume conditions are not violated: in particular, the latter condition is enforced by imposing that the destination lattice site is either empty or, at most, occupied by one and only one of the nearest-neighbor monomers along the chain.
In practice, it is useful to make the following conceptual distinction concerning how this move effectively implements classical polymer dynamics~\cite{DeGennes1979,DoiEdwardsBook,RubinsteinColbyBook}.
The selected monomer which is displaced towards an empty or a single-occupation site (or, a Rouse-like move) may, in general, change the local chain curvature and, hence, depend on the bending energy term Eq.~\eqref{eq:CosPotential}.
Conversely, a unit of stored length traveling along the chain (or, a reptation-like move) is not affected by the curvature because Eq.~\eqref{eq:CosPotential} is, again, strictly restricted to the effective bonds of the chains.
Overall, as explained in detail in~\cite{Schram-LatticeModel2018}, the stored length method ensures that the algorithm remains efficient even when it is applied to the equilibration of very dense systems.
The specific values for the acceptance rates as a function of the bending rigidity $\kappa_{\rm bend}$ are summarized in Table~\ref{tab:PolymerModel-LengthScales}.

\begin{table*}
\begin{tabular}{ccccccccccc}
\hline
\hline
\\
{\small $\kappa_{\rm bend}$} & {\small \mbox{acc. rate}} & {\small $\langle \cos\theta\rangle^{\rm lin}$} & {\small $\langle \cos\theta\rangle^{\rm ring}$} & {\small $\langle b \rangle / a$} & {\small $\ell_K / a$} & {\small $\rho_K \ell_K^3$} & {\small $L_e / a$} & {\small $N_e$} & {\small $d_T / a$} & {\small $\tau_e / \tau_{\rm MC} (\times 10^4)$} \\
\hline
$0$ & $0.069$ & $0.187$ & $0.171$ & $0.731$ & $1.440$ & $2.679$ & $58.738$ & $80.379$ & $3.755$ & $15.0$\\
$1$ & $0.048$ & $0.476$ & $0.447$ & $0.696$ & $2.194$ & $5.920$ & $20.708$ & $29.764$ & $2.752$ & $5.2$ \\
$2$ & $0.036$ & $0.670$ & $0.635$ & $0.669$ & $3.393$ & $13.620$ & $8.759$ & $13.088$ & $2.226$ & $2.8$ \\
\hline
\hline
\end{tabular}
\caption{
Values of physical parameters characterizing the melts of polymers studied in this paper.
$a$ is the unit distance of the fcc lattice and the monomer number per unit volume is $\rho a^3 \simeq 1.77$ (Sec.~3.2). 
(i) $\kappa_{\rm bend}$, bending stiffness parameter (Eq.~\eqref{eq:CosPotential});
(ii) acceptance rate of the kMC algorithm;
(iii, iv) $\langle \cos\theta\rangle^{\rm lin / ring}$, mean value of the cosine of the angle between two consecutive bonds along the linear/ring polymer (Eq.~\eqref{eq:CosPotential});
(v) $\langle b \rangle$, mean bond length;
(vi) $\ell_K$, Kuhn length (Eq.~\eqref{eq:Define-Lk}); 
(vii) $\rho_K \ell_K^3$, number of Kuhn segments per Kuhn volume~\cite{UchidaJCP2008};
(viii) $L_e$, entanglement length (Eq.~\eqref{eq:Uchida-etal});
(ix) $N_e \equiv L_e / \langle b\rangle$, number of bonds per entanglement length;
(x) $d_T = \sqrt{\ell_K L_e / 6}$, diameter of the effective tube confining polymer chains in melt;
(xi) $\tau_e$, entanglement time.
}
\label{tab:PolymerModel-LengthScales}
\end{table*}
%

\subsection{3.2 Melts of rings: simulation details\label{sec:SimulationDetails}}
We have studied bulk properties of dense solutions of $M$ closed (ring) polymer chains made of $N$ monomers or bonds per ring.
By construction, rings are {\it unknotted} and {\it non-concatenated}.
We simulate values
$$
N\times M = \left[ 40\times1000, 80\times500, 160\times250, 320\times125, 640\times62 \right] \, ,
$$
with fixed total number of monomers $=40,000$ (for computational convenience, the total number of monomers for $N=640$ is slightly less).
All these systems have been studied for the bending stiffness parameters $\kappa_{\rm bend} = 0,1,2$ (Eq.~\eqref{eq:CosPotential}).
In addition, we have also simulated melts with
$N\times M = \left[ 234 \times 171, 236 \times 58 \right]$ for $\kappa_{\rm bend}=1$
and
$N\times M = \left[ 104\times385, 104\times64 \right]$ for $\kappa_{\rm bend}=2$:
once rescaled in terms of the corresponding entanglement units (see Sec.~3.4 
and Table~\ref{tab:PolymerModel-LengthScales}), these two set-up's have the same number of entanglements per chain $Z\approx8$ of $N=640$ with $\kappa_{\rm bend}=0$.

Bulk conditions are implemented through the enforcement of periodic boundary conditions in a simulation box of total volume $V = L_{\rm box}^3$.
Similarly to previous works~\cite{Schram-LatticeModel2018,SchramRosaEveraers2019,ubertini2021computer}, melt conditions correspond to fix the monomer number per fcc lattice site to
(i) $\rho_{\rm site}=\frac54 = 1.25$ for $N\leq 320$
and
(ii) $\rho_{\rm site}=\frac{31}{25} = 1.24$ for $N= 640$:
respectively, since the volume occupied by the fcc lattice site is $=\frac{a^3}{\sqrt{2}}$~\cite{AshcroftMermin}, the monomer number per unit volume are given by
(i) $\rho = \frac54\sqrt{2} \simeq 1.77 a^{-3}$ for $N\leq 320$
and
(ii) $\rho = \frac{31}{25}\sqrt{2} \simeq 1.75 a^{-3}$ for $N= 640$.
Accordingly, we fix $L_{\rm box} = 20 \sqrt{2} a$ for all $N$'s.

Finally, our typical runs amount to a minimum of $7\times 10^6$ up to a maximum of $3\times 10^8$ kMC time units where one kMC time unit $\tau_{\rm MC}$ is $=NM$:
these runs are long enough such that the considered polymers have attained proper structural relaxation (see Fig.~S1 in SI). 

\subsection{3.3 A closer look to the bending stiffness\label{sec:Properties-Angles}}
It is worth discussing in more detail the consequences of the energy term, Eq.~\eqref{eq:CosPotential}. 

On the fcc lattice, the angle $\theta$ between consecutive chain bonds is restricted to the five values $= 0^\circ, 45^\circ, 90^\circ, 135^\circ, 180^\circ$. 
Otherwise, for ideal polymers ({\it i.e.} in the absence of the excluded volume interaction), the angles $\theta_i$ in Eq.~\eqref{eq:CosPotential} are obviously not correlated to each other. 
This implies that the distribution function $P_{\rm ideal}(\kappa_{\rm bend}; \cos \theta)$ of the variable $\cos\theta$ follows the simple Boltzmann form: 
\begin{equation}\label{eq:P_theta_ideal}
P_{\rm ideal} \left( \kappa_{\rm bend}; \cos \theta \right) = \frac1{\mathcal Z} \, {\mathcal N}\!(\cos\theta) \, e^{\kappa_{\rm bend} \cos\theta} \, ,
\end{equation}
where
${\mathcal N}\!(\cos\theta)$ represents the total number of lattice states for two successive bonds with given $\cos\theta$
and
$\mathcal Z$ is the normalization factor.

\begin{figure}
\includegraphics[width=0.46\textwidth]{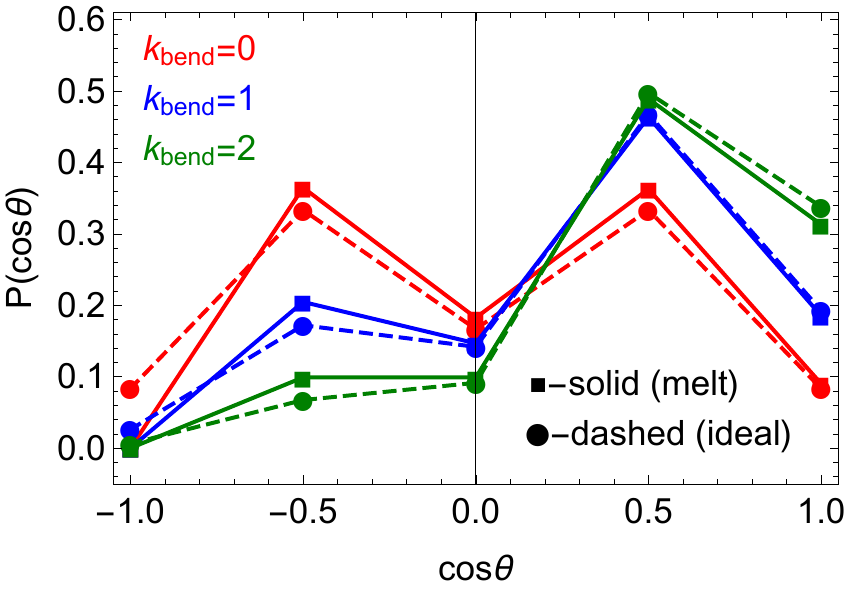}
\caption{\label{fig:P_theta}
Distribution functions $P_{\rm melt} \left( \kappa_{\rm bend}; \cos\theta \right)$ (square-solid lines) of the cosine of the angle $\theta$ between consecutive bonds along the polymer chain in ring melts
for different values of the bending stiffness parameter $\kappa_{\rm bend}$.
The curves are compared to the results for ideal rings, $P_{\rm ideal} \left( \kappa_{\rm bend}; \cos\theta \right)$ (Eq.~\eqref{eq:P_theta_ideal}, circle-dashed lines). 
}
\end{figure}

In polymer melts, excluded volume interactions induce an effective long range coupling between bond vectors and the distribution function $P_{\rm melt}(\kappa_{\rm bend}; \theta)$ is expected to deviate from Eq.~\eqref{eq:P_theta_ideal}:
for instance, one can immediately see that the angle $=180^\circ$ ({\it i.e.} $\cos\theta=-1$) is possible for ideal polymers but strictly forbidden in the presence of excluded volume interactions.
The distributions $P_{\rm melt}(\kappa_{\rm bend}; \theta)$ for the different $\kappa_{\rm bend}$'s 
and in comparison to $P_{\rm ideal}(\kappa_{\rm bend}; \theta)$ are given in Fig.~\ref{fig:P_theta} (square-solid {\it vs.} circle-dashed lines, respectively).
Corresponding mean values $\langle \cos\theta\rangle$ for the different $\kappa_{\rm bend}$'s and for open linear chains and closed rings are summarized in Table~\ref{tab:PolymerModel-LengthScales}:
the values for the two chain architectures are very similar, with differences in the range of a few percent.

\subsection{3.4 Polymer length and time scales\label{sec:Properties-PolymerSols}}
We provide here a detailed derivation of the relevant length and time scales characterizing our polymer melts (they are summarized in Table~\ref{tab:PolymerModel-LengthScales}).
Since we work at fixed monomer density $\rho$ (see Sec.~3.2), 
the values of these parameters depend only on the bending stiffness parameter $\kappa_{\rm bend}$.

{\it Average bond length}, $\langle b\rangle$ -- 
We observe that $\langle b \rangle$ is slightly decreasing with $\kappa_{\rm bend}$:
arguably, this is the consequence of the progressive stiffening of the polymer fiber on the fcc lattice which privilege less kinked conformations through the reduction of the effective total contour length of the chain.

\begin{figure}
\includegraphics[width=0.47\textwidth]{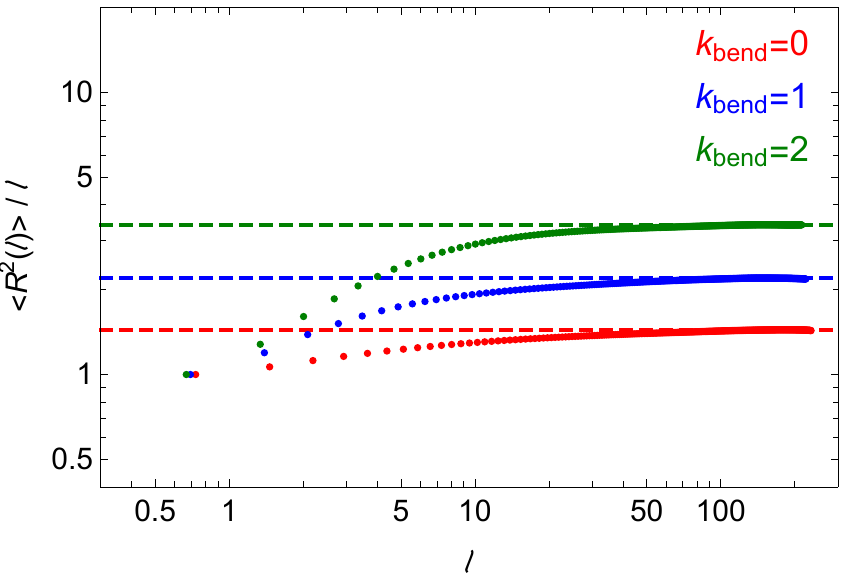}
\caption{\label{fig:lk}
(Symbols)
Ratios of the mean-square end-to-end distance, $\langle R^2(\ell) \rangle / \ell$, to the corresponding monomer contour length separation $\ell$ for linear chains in melt and for the different bending stiffness parameters $\kappa_{\rm bend}$ (see legend).
(Dashed lines)
Best fits of the plateau-like behaviors on the contour length interval $\ell\in[150,200]$ (see Eq.~\eqref{eq:Define-Lk} and the text for details).
}
\end{figure}

{\it Kuhn length}, $\ell_K$ --
By modulating the bending constant $\kappa_{\rm bend}$, we can fine-tune the flexibility of the polymers (Eq.~\eqref{eq:CosPotential}).
The latter is quantified in terms of the {\it Kuhn length} $\ell_K$, namely the unit of contour length beyond which the chain orientational order is lost~\cite{DoiEdwardsBook,RubinsteinColbyBook}. 
For {\it linear chains}, $\ell_K$ is defined by the relation~\cite{DoiEdwardsBook,RubinsteinColbyBook}:
\begin{equation}\label{eq:Define-Lk}
\ell_K \equiv \lim_{\ell\rightarrow\infty} \frac{\langle R^2(\ell) \rangle}{\ell} \, ,
\end{equation}
where $\langle R^2(\ell) \rangle$ is the mean-square end-to-end distance between any two monomers along the chain at monomer number separation $n$ or contour length separation $\ell = n\langle b \rangle$. 
In order to determine the Kuhn length of our polymer chains, 
we have simulated melts of $M=125$ linear chains with $N=320$ monomers per chain at the same monomer density $\rho = \frac54\sqrt{2} \simeq 1.77 a^{-3}$ (Sec.~3.2) 
and for $\kappa_{\rm bend}=0,1,2$ and, after equilibration, we have computed numerically Eq.~\eqref{eq:Define-Lk}. 
As shown in Fig.~\ref{fig:lk} (symbols), the chains become increasingly stiffer with $\kappa_{\rm bend}$ as expected by displaying, in particular, plateau-like regions for large $\ell$'s.
In analogy to the procedure employed in~\cite{ubertini2021computer}, the heights of these plateaus, obtained by best fits to corresponding constants on the common interval $\ell \in [150, 200]$ (Fig.~\ref{fig:lk}, dashed lines), provide the values of the corresponding $\ell_K$'s which are used in this work (see Table~\ref{tab:PolymerModel-LengthScales}).

{\it Entanglement length}, $L_e$, and {\it tube diameter}, $d_T$ --
The entanglement length $L_e$ marks the crossover from entanglement-free to entanglement-dominated effects in polymer melts. In general $L_e$ depends in a non-trivial~\cite{Lin1987,KavassalisNoolandiPRL1987} manner on the microscopic details of the polymer melt, typically the chain Kuhn length $\ell_K$ and the monomer density $\rho$.
By combining packing arguments~\cite{MorsePRE98} and primitive path analysis~\cite{EveraersScience2004}, Uchida {\it et al.}~\cite{UchidaJCP2008} showed that $L_e$ can be expressed as a function of $\rho_K \ell_K^3$, the number of Kuhn segments in a Kuhn volume at given Kuhn density $\rho_K \equiv  \frac{\rho}{\ell_K/a}$:
\begin{equation}\label{eq:Uchida-etal}
\frac{L_e}{\ell_K} =  \left(\frac1{0.06 \, \rho_K \ell_K^{3}}\right)^{2/5} + \left(\frac1{0.06 \, \rho_K \ell_K^{3}}\right)^2 \, ,
\end{equation}
where the two power laws account for the two limits of {\it loosely} entangled melts ($\rho_K \ell_K^3 \ll 1$, exponent$=2$) and {\it tightly} entangled solutions ($\rho_K \ell_K^3 \gg 1$, exponent$=2/5$).
By using\footnote{Strictly speaking, Eq.~\eqref{eq:Uchida-etal} was derived for the entanglement length of a melt of {\it linear} chains. Yet, recent theoretical work~\cite{HalversonPRL2012,RosaEveraersPRL2014,SchramRosaEveraers2019} on {\it ring} melts has demonstrated that {\it the same} $L_e$ describes the correct contour length where topological constraints affect ring behavior and chains become manifestly crumpled.} Eq.~\eqref{eq:Uchida-etal}, it is a simple exercise to extract $L_e / \ell_K$ and the corresponding number of monomers per entanglement length, $N_e \equiv L_e / \langle b \rangle$. 
We observe (Table~\ref{tab:PolymerModel-LengthScales}) that $N_e$ decreases by approximately one order of magnitude by the apt fine-tuning of $\kappa_{\rm bend}$. 
This means that rings with the same contour length but stiffer will become increasingly entangled~\cite{Cates2000}. 
Finally, we use the definition Eq.~\eqref{eq:DefineTubeDiameter}:
$$
d_T = \sqrt{\frac{L_e \ell_K}6} \, ,
$$
for computing the tube diameters of polymers of different bending stiffnesses.
Notice (Table~\ref{tab:PolymerModel-LengthScales}) that, by changing chain stiffness, $d_T$ moves from smaller to comparable to $\ell_K$ meaning that we are effectively able to explore the mentioned crossover from loosely to tightly entangled melts. 

{\it Entanglement time}, $\tau_e$ --
The entanglement time $\tau_e$ marks the onset to entanglement-related effects in polymer dynamics.
As anticipated in Sec.~2.1, $\tau_e$ is evaluated numerically as $g_1(\tau_e) = 2d_T^2$, where  $g_1(\tau)$ is the monomer time mean-square displacement (Eq.~\eqref{eq:Introduce-g1}) and $d_T$ is the tube diameter (Eq.~\eqref{eq:DefineTubeDiameter}).
For consistency with the definitions for $L_e$ and $d_T$, notice that $g_1(\tau)$ has been calculated on the same dynamic simulations of melts of {\it linear} chains used for calculating $\ell_K$.

\subsection{3.5 Computing the ring minimal surface\label{sec:ComputingMinS}}
The minimal surface spanned by a ring polymer was introduced~\cite{LangMacromolecules2013,SmrekGrosbergACSMacroLett2016} as a quantitative tool for measuring the ``exposed'' area that each ring offers to its neighbors.
In this section, we limit ourselves to summarize the salient numerical aspects of the procedure used to obtain the minimal surface, the interested reader will find a complete overview on minimal surfaces for melts of rings in the mentioned references and also in~\cite{SmrekRosa2019}.

The search for the minimal surface of a ring polymer is based on a suitable minimization algorithm which works as the following.
Essentially, the algorithm is based on successive iterations of triangulations evolving under surface tension by moving the free vertices:
each triangle in the initial triangulation is made of two successive monomer positions and the center of the mass of the ring, then it is refined (by subdividing each edge into two edges, creating therefore four triangles out of one) and the surface area minimized by the surface tension flow with restructuring of the mesh.
Finally, the algorithm stops when the relative surface area does not change by more than $0.1\%$ over a few tens of additional steps of the minimization procedure.
Similarly to the eye of a needle pierced by a thread, the single minimal surface of a given ring can be pierced or {\it threaded} by the other surrounding polymers in the melt:
in particular, once the minimal surfaces of the rings in the melt are determined,
it is possible to define in a precise and robust manner what amount of the total contour length of any given ring passes through the minimal surface of another ring.

\section{4. Results\label{sec:Results}}

\subsection{4.1 Ring structure\label{sec:UntangledRings-Structure}}

%
\begin{figure}
\includegraphics[width=0.45\textwidth]{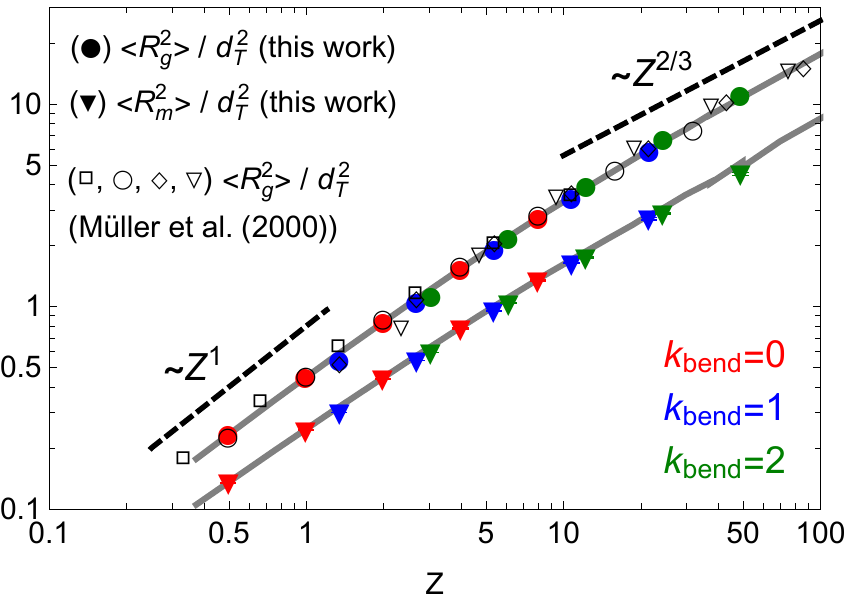}
\caption{\label{fig:RgUntangledRings}
Scaling behavior of the mean-square gyration radius, $\langle R_g^2\rangle$, and the mean-square magnetic radius, $\langle R_m^2\rangle$, of ring polymers as a function of the total number of entanglements, $Z=L/L_e$, of the chain.
Error bars are smaller than the symbols size.
The empty symbols reproduce the ring data by M\"uller {\it et al.}~\cite{Cates2000}, with different symbols for the different Kuhn lengths used in that work (by employing their notation: $\sigma/\kappa_BT=0(\square),1(\circ),2(\diamond),3(\triangledown)$, from flexible to increasingly stiffer chains).
The solid grey lines are for the on-lattice rings on the fcc lattice studied in~\cite{SchramRosaEveraers2019}.
}
\end{figure}

In order to characterize ring structure, we consider the chain mean-square ``gyration'' radius,
\begin{equation}\label{eq:Rg2}
\langle R_g^2 \rangle = \left\langle \frac1N \sum_{i=1}^N ( \vec r_i - \vec r_{\rm cm} )^2 \right\rangle \, ,
\end{equation}
where $\vec r_i$ are the monomer coordinates and $\vec r_{\rm cm} = \frac1N \sum_{i=1}^N \vec r_i$ is the chain centre of mass,
and the mean-square ``magnetic'' radius first introduced in~\cite{SchramRosaEveraers2019} and defined as
\begin{equation}\label{eq:MagnRadius}
\langle R_m^2 \rangle = \frac1\pi \langle \left| \vec A \right| \rangle \, ,
\end{equation}
where, inspired by the analogy to the classical electrodynamics of the magnetic far field generated by a loop carrying a constant electric current,
\begin{equation}\label{eq:EnclosedArea}
\vec A \equiv \frac12 \sum_{i=1}^N \vec r_i \times (\vec r_{i+1}-\vec r_i) = \frac12 \sum_{i=1}^N \vec r_i \times \vec r_{i+1} 
\end{equation}
is the (oriented) area enclosed by the ring.
Both quantities are an expression of the average (square) ring size, yet they have different meaning: in particular, Eq.~\eqref{eq:MagnRadius} was introduced to detect and quantify the presence of open loops inside the ring.
Numerical values of $\langle R_g^2 \rangle$ and $\langle R_m^2 \rangle$ for melts of rings with different flexibilities are summarized in Table~S1 in SI. 

Fig.~\ref{fig:RgUntangledRings} shows these quantities, rescaled (filled symbols) by the corresponding tube diameters $d_T$ and as a function of the total number of entanglement $Z=L / L_e$, see Sec.~2.1 
for details and Table~\ref{tab:PolymerModel-LengthScales} for specific values of $d_T$ and $L_e$.
For further comparison, we have also included the results for $\langle R_g^2\rangle$ obtained from Monte Carlo simulations (bond-fluctuation model) of melts of rings by M\"uller {\it et al.}~\cite{Cates2000} (open symbols) and the universal curves (solid grey lines) from the ``hierarchical crumpling'' method by Schram {\it et al.}~\cite{SchramRosaEveraers2019}.
The excellent matching between these old data sets and the present new data validates our methodology:
the average ring size covers the full crossover from Gaussian ($\sim Z^1$) to compact ($\sim Z^{2/3}$) behavior.
This is particularly evident for the stiffest rings ($\kappa_{\rm bend}=2$) whose reduced flexibility ``helps'', in agreement with M\"uller {\it et al.}~\cite{Cates2000}, reaching the asymptotic behavior. 
The data for the mean-square magnetic radius $\langle R_m^2 \rangle$ rescale equally well and, as noticed in~\cite{SchramRosaEveraers2019}, with the same large-$Z$ behavior $\sqrt{ \langle R_m^2 \rangle} \sim d_T \, Z^{1/3}$.
However a closer look to the ratio $\langle R_g^2 \rangle / \langle R_m^2 \rangle$ (see Fig.~S2 in SI) 
reveals significant, albeit extremely slow (${\mathcal O}(Z^\gamma)$ with $\gamma \approx 0.07$), power-law corrections which {\it were not} noticed in the previous study~\cite{SchramRosaEveraers2019}:
instead, a similar result was already described in~\cite{Cates2000} where, however, the authors employed a different definition for the area spanned by a ring which was based on the $2d$ projection of the chain onto a random direction.

\begin{figure}
\includegraphics[width=0.47\textwidth]{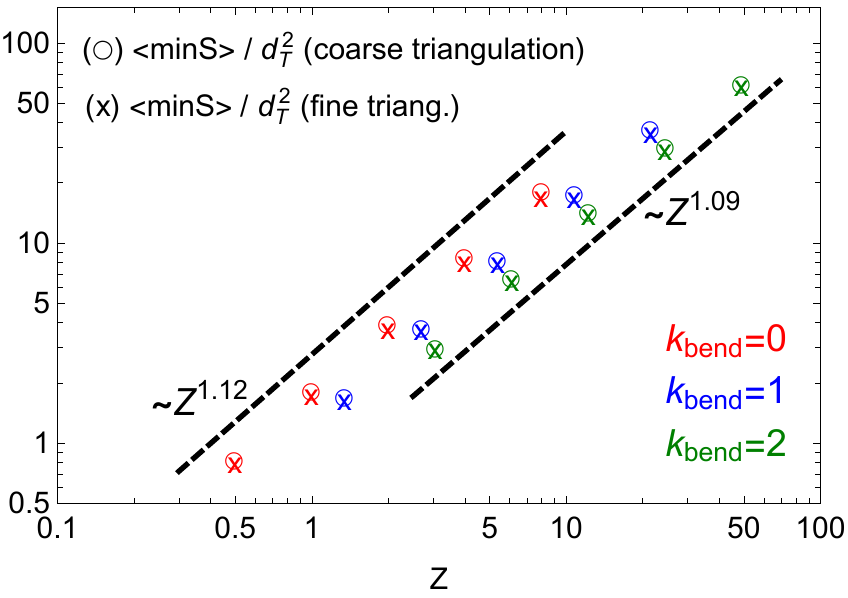}
\caption{\label{fig:minSvsN}
Mean minimal surface area, $\langle {\rm minS} \rangle$, of ring polymers as a function of the total number of entanglements, $Z$, of the chain.
The $\circ$- and $\times$-symbols (see legend) are for the two chosen resolutions in the triangulation procedure at the basis of the calculation of the ring minimal surface (see Sec.~3.5 for details).
For the higher resolution, we take two additional refining steps before the minimization procedure stops.
Error bars are smaller than the symbols size.
The power laws (dashed lines) correspond to best fits to the data.
}
\end{figure}

The universal behaviors of the two quantities $\langle R_g^2\rangle$ and $\langle R_m^2\rangle$ characterize the chain as a whole yet, in principle, there could be other length scales below $L_e$ which produce no effect on these quantities but affect and become visible in others.
We look then at the average behavior of the ring minimal surface (hereafter, $\rm minS$), a concept introduced~\cite{LangMacromolecules2013,SmrekGrosbergACSMacroLett2016,SmrekRosa2019} to quantify the ``exposed'' area that each ring offers to its neighbors (see Sec.~3.5 for technical details).
Surprisingly, our data (collected in Table~S1 in SI) demonstrate that this is not the case (see Fig.~\ref{fig:minSvsN}):
in fact $\langle {\rm minS}\rangle \sim Z$ for the different flexibilities in agreement with previous results~\cite{SmrekGrosbergACSMacroLett2016} for off-lattice simulations, but the three data sets -- contrary to $\langle R_g^2\rangle$ and $\langle R_m^2\rangle$ (Fig.~\ref{fig:RgUntangledRings}) -- do not collapse onto each other after rescaling in terms of entanglement units $Z$ and $d_T^2$.
Notice that this finding -- which looks even more surprising given that both, $\langle R_m^2\rangle$ and $\langle {\rm minS}\rangle$, are in the end two measures of the ring effective area -- is robust and independent of the triangulation resolution employed to calculate the minimal surface of the rings (``$\circ$-symbols'' vs. ``$\times$-symbols'' in Fig.~\ref{fig:minSvsN}).
Note also that $\langle {\rm minS}\rangle$ is normalized by $d_T^2$ which should correct for the different ``elementary areas'' due to different local structure of the polymer model\footnote{Note also that normalization by the other microscopic length scale, $\ell_K$, would not imply a collapse either, in fact it would make it even worse as $\ell_K$ grows with $\kappa_{\rm bend}$ (see Table~\ref{tab:PolymerModel-LengthScales}).}.

\begin{figure}
\includegraphics[width=1.00\textwidth]{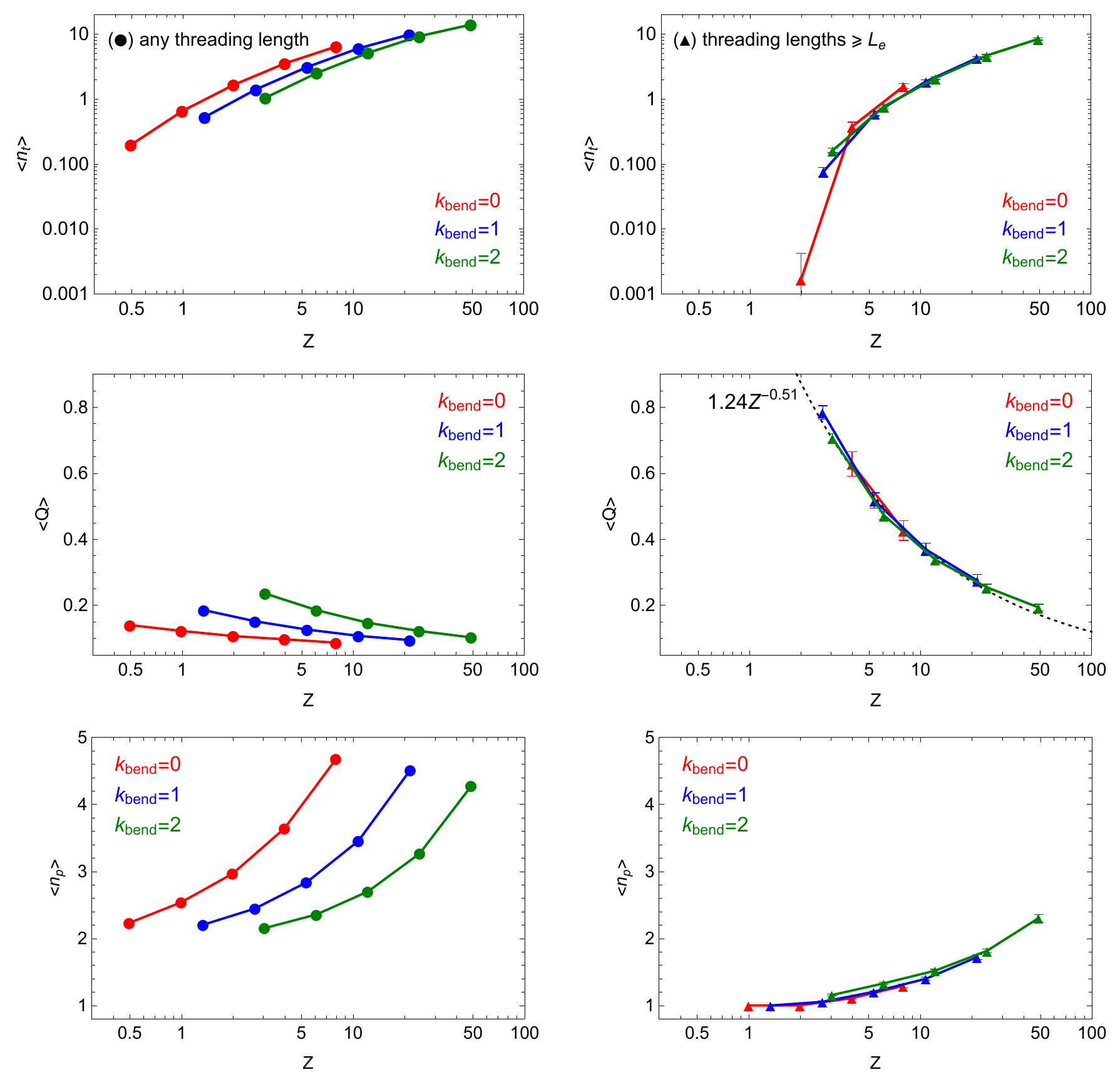}
\caption{\label{fig:ntvsZ}
Statistics of threadings by minimal surfaces as a function of the total number of entanglements, $Z$, of each ring in the melt.
The data (circles) shown in the l.h.s panels (respectively, (up-triangles) in the r.h.s panels) have been obtained by including threading lengths $L_t$ of any size (resp., $L_t \geq L_e$).
Error bars represent the error of the mean calculated from $100$ uncorrelated snapshots.
(Top)
$\langle n_t \rangle$: mean number of rings threaded by a single ring. 
(Middle)
$\langle Q \rangle$: mean relative amount of contour length piercing one ring minimal surface side with respect to the other (Eq.~\eqref{eq:separation_length-Q}). 
The dashed (black) line on the r.h.s. panel corresponds to the best fit to the data.
(Bottom)
$\langle n_p \rangle$: mean number of times a ring penetrates the minimal surface of any other single ring. 
}
\end{figure}

What could be the physical origin of this discrepancy?
In an earlier paper~\cite{SmrekRosa2019} it was shown that the minimal surface of each ring can be pierced (threaded) a certain number of times by portions of surrounding rings, but that it is necessary to distinguish carefully the contributions from chain contour lengths $z = \ell / L_e \lesssim 1$ ({\it shallow threadings}) from the others.
For each ring piercing the minimal surface of another ring, we define~\cite{SmrekRosa2019} the threading contour length $L_{t_i}$ as the contour length portion of the ring comprised between the two consecutive penetrations $i$-th and $i+1$-th which allow us to assign $L_{t_i}$ {\it unambiguously} to either ``side'' of the other ring's minimal surface.
Based on $L_{t_i}$, we consider the following observables:
\begin{enumerate}
\item
The mean number of chain neighbors threaded by a single ring, $\langle n_t\rangle$.
\item
The mean relative amount of contour length of a ring on one side of another ring's minimal surface with respect to the other side:
\begin{equation}\label{eq:separation_length-Q}
\langle Q \rangle \equiv \left\langle \frac{L_{\rm sep}}{L-L_{\rm sep}} \right\rangle \, ,
\end{equation}
where the so called {\it separation length}~\cite{SmrekGrosbergACSMacroLett2016,SmrekRosa2019},
\begin{equation}\label{eq:separation_length}
L_{\rm sep} = \min_{}\left(\sum_{i=\mathrm{even}}L_{t_{i}},\sum_{i=\mathrm{odd}}L_{t_{i}}\right) \, ,
\end{equation}
quantifies for the total contour length of one of the two complementary portions (the other has contour length $L-L_{\rm sep}$) obtained by the passage of the {\it penetrating} ring through the minimal surface of a {\it penetrated} ring:
{\it i.e.}, $Q=1$ simply means that the penetrating ring is half split by the penetrated surface.
\item
The mean number of times any ring penetrates the minimal surface of any other ring, $\langle n_p \rangle$.
\end{enumerate}
These different quantities may be computed by taking into account all possible threading segments or by excluding the shallow ones namely, as said, those whose contour lengths are shorter than the entanglement length $L_e$. Noticeably, in the first case there is no evidence for universal collapse (Fig.~\ref{fig:ntvsZ}, l.h.s. panels) in any of these quantities.
On the contrary (Fig.~\ref{fig:ntvsZ}, r.h.s. panels), universal collapse is observed after removing the contribution of the short threading filaments: this points to the fact that the same shallow threadings are responsible for the lack of universality observed for $\langle {\rm minS}\rangle$ (Fig.~\ref{fig:minSvsN}).
Assuming that each shallow threading contributes $\sim d_T^2$ to the average $\langle {\rm minS} \rangle$, their removal restores universality almost completely (Fig.~S3 in SI). To justify the assumption that each threading does not contribute more than about $\sim d_T^2$ to the minimal surface, we measured the mean distance $\langle d \rangle$ between two bonds involved in the threading.
We find that indeed $\langle d \rangle \lesssim d_T$ (see the values, the corresponding distribution functions as well as the discussion of the cases with more than two piercings in Fig.~S4 in SI), which strongly supports the fact that threadings occur on scales below the entanglement scale.
We still find some cases when $\langle d \rangle > d_T$, but these are exponentially rare (see Fig.~S4 in SI).

Additionally, restricting the discussion to the universal behaviors, we see that $\langle Q\rangle$ decreases with $Z$, {\it i.e.} less material enters the minimal surface of a ring.
Yet, the mean number of times, $\langle n_p \rangle$, any ring penetrates the minimal surface of any other single ring increases: notice that these two apparent contradictory features can be easily reconciled by supposing that while retracting from each other ({\it i.e.} decreasing $\langle Q\rangle$), rings may at the same time pierce each other more frequently by the increased propensity to form branches at the entanglement scale.
Overall, this agrees with the results by two of us (J.S., A.R.)~\cite{SmrekRosa2019} for off-lattice dynamic simulations of melts of rings\footnote{Notice, however, that in~\cite{SmrekRosa2019} we have reported the scaling behavior $\langle Q\rangle \sim Z^{-0.31}$, distinct from the one measured here ($\langle Q\rangle \sim Z^{-0.51}$, see middle right panel in Fig.~\ref{fig:ntvsZ}). This apparent discrepancy is due to the fact that in~\cite{SmrekRosa2019} the mean value $\langle Q\rangle$ includes the contribution of the shallow threadings: in fact, their removal gives identical results to the ones of the present work (data not shown).}.

According to the lattice-tree model (Sec.~2.2), the contour length of each ring above the entanglement length $L_e$ should double-fold along a branched ({\it i.e.}, tree-like) backbone. We seek now specific evidences of this peculiar organization.
As the first of these signatures, we look at the bond-vector correlation function~\cite{Cates2000,RosaEveraersPRL2014,SchramRosaEveraers2019}
\begin{equation}\label{eq:tgtgCorrel}
c(\ell) \equiv \frac{\langle \vec t(\ell') \cdot \vec t(\ell'+\ell) \rangle}{\langle \vec t(\ell')^2 \rangle} \, .
\end{equation}
\begin{figure}
\includegraphics[width=0.45\textwidth]{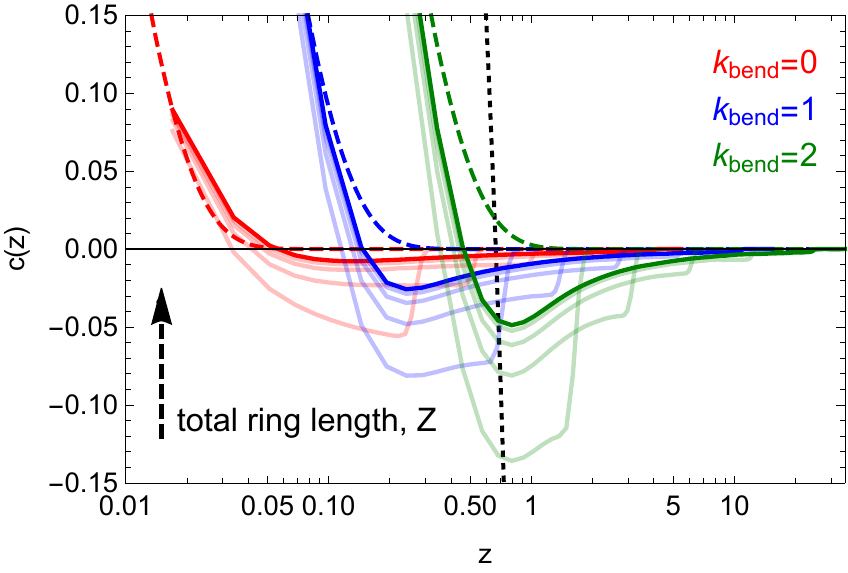}
\caption{\label{fig:tgtgCorrel}
Bond-vector orientation correlation function, $c(z)$ (Eq.~\eqref{eq:tgtgCorrel}), as a function of the number of entanglements $z=\ell/L_e$ along the chain.
Lines of equal color are for the same chain stiffness (see legend), full colors are for the longest rings ($N=640$) while lines in fainter colors are for chains of shorter contour lengths (see arrow's direction).
The long-dashed lines correspond to the exponential decay typical of linear polymers with local stiffness, namely $c(\ell) = (\langle\cos\theta\rangle^{\rm ring})^{\ell/\langle b \rangle}$, where the values for $\langle\cos\theta\rangle^{\rm ring}$ are reported in Table~\ref{tab:PolymerModel-LengthScales}.
The short-dashed line is the analytical function for an exactly double-folded polymer filament of contour length $=2L_e$ (see the text for detail).
}
\end{figure}
Contrary to the known exponentially-decaying behavior typical of linear polymers (solid lines {\it vs.} long-dashed lines in 
Fig.~\ref{fig:tgtgCorrel}) $c(\ell)$ is manifestly non-monotonic, showing an anti-correlation well whose minimum becomes more pronounced and locates around $z = \ell/L_e \approx 1$ with the increasing of the chain stiffness.
By using the function $\vec t(\ell) = (0,+1,0)$ for $0 < \ell < L_e$ and $\vec t(\ell) = (0,-1,0)$ for $L_e < \ell < 2L_e$ as a toy model for an exactly double-folded polymer filament of contour length $=2L_e$, it is easy to see that $c(z=\ell/L_e) = (1-3/2z) / (1-z/2)$ for $0<z<1$ and $c(z)=-1$ for $1 \leq z<2$, {\it i.e.} $c(\ell)$ displays an anti-correlation minimum at $z=1$ or $\ell=L_e$ (short-dashed line in 
Fig.~\ref{fig:tgtgCorrel}).
A deep anti-correlation well is particularly visible in short rings (at given $\kappa_{\rm bend}$) while in larger rings the effect is smoothed, arguably because thermal fluctuations wash out such strong anti-correlations.
As a second signature, we look at the mean contact probability~\cite{Dekker_HiC2009}, $\langle p_c(\ell) \rangle$, between two monomers at given contour length separation $\ell$ defined as:
\begin{equation}\label{eq:DefinePc}
\langle p_c(\ell \equiv n\langle b \rangle) \rangle = \left\langle \frac2{N(N-1)} \sum_{i=1}^{N-1} \sum_{j=i+1}^N \Theta(r_c-|\vec r_i - \vec r_j|)\right\rangle \, ,
\end{equation}
where $\Theta(x)$ is the usual Heaviside step function and the ``contact distance'' $r_c$ between two monomers is taken equal to one lattice unit, $a$.
\begin{figure}
\includegraphics[width=0.45\textwidth]{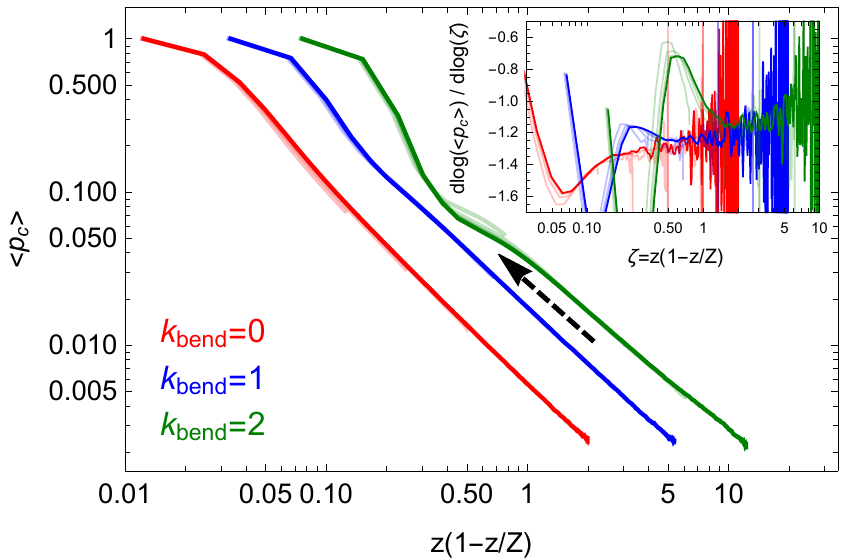}
\caption{\label{fig:pc}
Mean contact probabilities, $\langle p_c(\ell) \rangle$ (see Eq.~\eqref{eq:DefinePc}), as a function of the ``effective'' ring contour length $\zeta \equiv z(1-z/Z)$.
The arrow points at the levelling of the curves around $z\approx 1$.
Inset: local differential exponent $\equiv \frac{d\log\langle p_c\rangle}{d\log\zeta}$.
Color code is as in Fig.~\ref{fig:tgtgCorrel}. 
}
\end{figure}
Reducing finite ring size effects by plotting $\langle p_c \rangle$ in terms of the variable~\cite{rosa2019conformational} $\zeta \equiv z(1-z/Z)$, the data from the different rings (see Fig.~S5 in SI) 
form three distinct curves according to their Kuhn length (Fig.~\ref{fig:pc}). 
Notably these curves display the asymptotic power-law behavior $\sim \zeta^{-\gamma}$ with scaling exponent $\gamma$ close to $1$, as reported in the past~\cite{halverson2011molecular-statics,HalversonSmrekRPP_2014,RosaEveraersPRL2014}. 
At the same time, the stiffer chains with $\kappa_{\rm bend}=2$ display a short, yet quite evident, levelling of the contact probability curves around $z\approx 1$ (see arrow's direction)
which is clearly compatible with double-folding on the entanglement scale.
\begin{figure}
\includegraphics[width=0.45\textwidth]{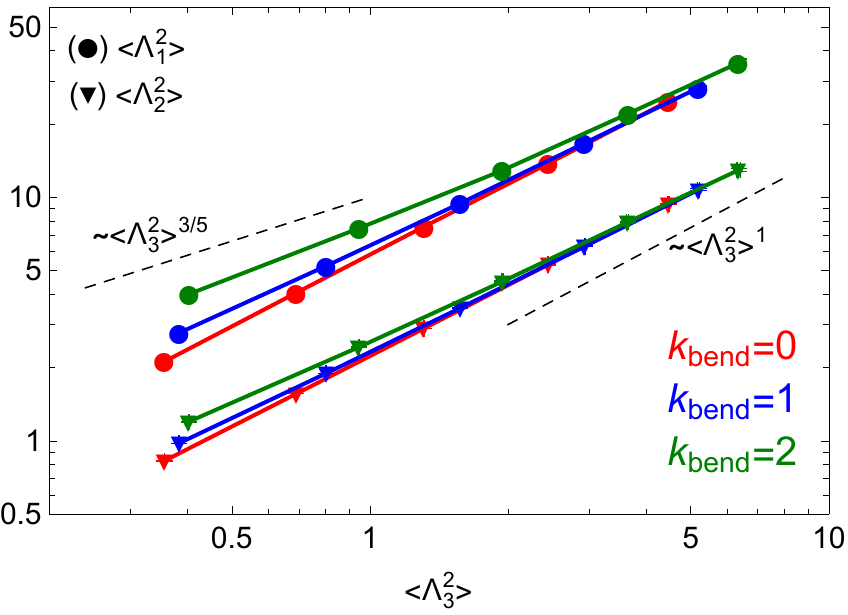}
\caption{\label{fig:LambdaSq1vsLambdaSq3etc}
Mean eigenvalues, $\langle \Lambda_1^2 \rangle$ and $\langle \Lambda_2^2 \rangle$, of the gyration tensor Eq.~\eqref{eq:Gyration_tensor} as a function of the smallest eigenvalue $\langle\Lambda_3^2 \rangle$.
The dashed lines are the predicted limits of small and large length scales.
}
\end{figure}
However, we show now that the most compelling evidence for double-folding comes from examining the average polymer {\it shape}.
To this purpose,
we introduce the $3\times 3$ symmetric gyration tensor $\mathcal G$ whose elements are defined by:
\begin{equation}\label{eq:Gyration_tensor}
{\mathcal G}_{\alpha\beta} = \frac{1}{2N^2} \sum_{i=1}^N \sum_{j=1}^N \left( r_{i, \alpha} - r_{j, \alpha} \right) \left( r_{i,\beta} - r_{j, \beta} \right) \, ,
\end{equation}
where $r_{i, \alpha}$ with $\alpha=x,y,z$ are the Cartesian components of the spatial position of monomer $i$:
the trace of the tensor $\mathcal G$, ${\rm Tr}{\mathcal G} \equiv \sum_{\alpha=1}^3 \Lambda_{\alpha}^2$, is equal to $R_g^2$ and the three ordered eigenvalues, $\Lambda_1^2 \geq \Lambda_2^2 \geq \Lambda_3^2$, quantify the spatial variations of the polymer along the corresponding principal axes, {\it i.e.} the instantaneous shape of the chain.
Polymers are ellipsoidal on average~\cite{BishopMichels1986}, with mean values $\langle \Lambda_1^2 \rangle > \langle \Lambda_2^2 \rangle > \langle \Lambda_3^2 \rangle$ (see values in Table~S1 in SI): 
similarly to $\langle R_g^2\rangle$ (Fig.~\ref{fig:RgUntangledRings}), we do expect a scaling curve for each $\langle \Lambda_\alpha^2 \rangle$ in universal units $Z$ and $d_T$ and, for $Z\gg 1$, $\langle \Lambda^2_{\alpha=1,2,3} \rangle \sim \langle R_g^2 \rangle \sim d_T^2 Z^{2\nu}$.
In general our data (see Fig.~S6 in SI) 
reflect well this trend,
except for the smallest mean eigenvalue $\langle \Lambda^2_3 \rangle$ which displays systematic deviations from the asymptotic behavior which persist for chain sizes well above the entanglement threshold $Z \approx 1$.
Interestingly, these deviations are {\it quantitatively} consistent with the scaling properties of rings being double-folded on an underlying tree-like structure (Sec.~2.2): 
in fact, according to Fig.~S6 in SI 
and for the given mean path length $\langle L_{\rm tree} \rangle \sim Z^{\rho}$ (Eq.~\eqref{eq:Rings-LvsZ}) of the ``supporting'' tree, we can write the expressions:
\begin{eqnarray}
\langle \Lambda_{i=1,2}^2 \rangle & \sim & Z^{2\nu} \sim \langle L_{\rm tree} \rangle^{2\nu_{\rm path}} \, , \label{eq:Lambda2-crossovers-1} \\
\langle \Lambda_3^2 \rangle & \sim & Z^{2\rho} \sim \langle L_{\rm tree} \rangle^2 \, , \label{eq:Lambda2-crossovers-2}
\end{eqnarray}
where the latter is equivalent to assuming local stiff behavior at small contour length separations.
Eqs.~\eqref{eq:Lambda2-crossovers-1} and~\eqref{eq:Lambda2-crossovers-2} are just equivalent to:
\begin{equation}\label{eq:Lambda2-crossovers-3}
\langle \Lambda_{i=1,2}^2 \rangle \sim \langle \Lambda_3^2 \rangle^{\nu_{\rm path}} \, ,
\end{equation}
with $\nu_{\rm path}=3/5$ (Eq.~\eqref{eq:Rings-RvsL}).
Our data are well described by Eq.~\eqref{eq:Lambda2-crossovers-3}, see Fig.~\ref{fig:LambdaSq1vsLambdaSq3etc}, before the crossover to the asymptotic regime, $\langle \Lambda_{i=1,2}^2 \rangle \sim \langle \Lambda_3^2 \rangle$, takes finally place.
Notice that the value of $\nu_{\rm path}$ agrees well with that of the annealed tree model~\cite{GrosbergSM2014}.
Note also that here we observe the exponent clearly only for the stiffest system.
This might explain why the work~\cite{PanyukovRubinsteinMacromolecules2016} that analyzed only simulations~\cite{halverson2011molecular-statics} with low stiffness (roughly, between our $\kappa_{\rm bend}=0$ and $\kappa_{\rm bend}=1$ in terms of $L_e$) reports that they do not see $\nu_{\rm path}$ (there measured as the scaling of the chain {\it primitive path} with the contour length).

In conclusion, the {\it universal} scaling features of the static quantities and the bond correlations reveal double-folded tree-like structure on scales above $L_e$, with indications ($\langle \Lambda_3^2 \rangle$) that the trees are of the annealed tree type~\cite{GrosbergSM2014}.
Conversely, the {\it non-universal} features ($\langle {\rm minS}\rangle$) and threading analysis show that the trees are threaded on scales smaller than the tube diameter $d_T$, hence reconciling the tree picture with that of the threaded conformations.
At the same time the ``depth'' of threadings can be (Fig.~\ref{fig:ntvsZ}) longer than $L_e$, giving hints on its relevance for the dynamic features.
Therefore, in the next section we explore the consequences of the double-folding and threadings on the dynamic properties of the melts.

\subsection{4.2 Ring dynamics\label{sec:UntangledRings-Dynamics}}

%
\begin{figure*}
\includegraphics[width=1.0\textwidth]{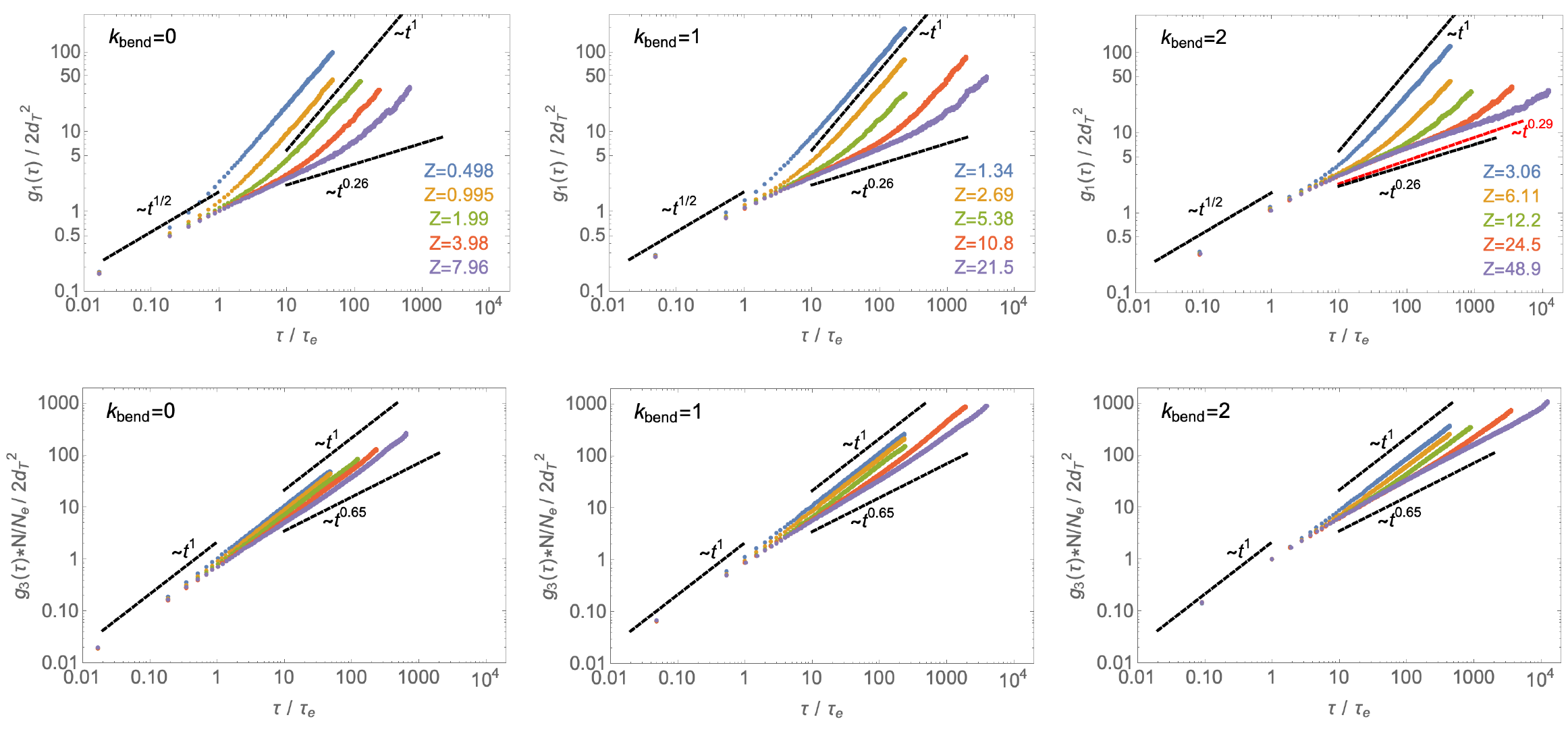}
\caption{\label{fig:g1g3}
Monomer time mean-square displacement, $g_1(\tau)$ (Eq.~\eqref{eq:Introduce-g1}), and chain center of mass time mean-square displacement, $g_3(\tau)$ (Eq.~\eqref{eq:Introduce-g3}).
Symbols of the same color are for the same number of monomers $N$, and the corresponding number of entanglements $Z$ per chain is indicated in the legends.
The dashed black lines correspond to the dynamic predictions of the lattice tree model, Eqs.~\eqref{eq:LatticeTree-g1-Complete} and~\eqref{eq:LatticeTree-g3-Complete} with $\nu=1/3$ (Eq.~\eqref{eq:Rings-RvsZ}) and $\rho=5/9$ (Eq.~\eqref{eq:Rings-LvsZ}).
The dashed red line in the top-right panel corresponds to the theoretical prediction of the FLG model~\cite{PanyukovRubinsteinMacromolecules2016} by Ge {\it et al.}.
}
\end{figure*}

As briefly discussed in Sec.~2.3, 
the lattice-tree model predicts universal dynamic chain behavior for length scales $\gtrsim d_T$ and time scales $\gtrsim \tau_e$.
Recent numerical works employing off-lattice molecular dynamics simulations~\cite{halverson2011molecular-dynamics} and lattice models~\cite{ElhamPRE2021} agree well with these predictions. In order to compare the dynamic behavior of the present systems to the results of the lattice-tree model (namely, Eqs.~\eqref{eq:LatticeTree-g1-Complete} and~\eqref{eq:LatticeTree-g3-Complete} with $\nu=1/3$ (Eq.~\eqref{eq:Rings-RvsZ}) and $\rho=5/9$ (Eq.~\eqref{eq:Rings-LvsZ})), we look then at the monomer mean-square displacement, $g_1(\tau)$, and the center of mass mean-square displacement, $g_3(\tau)$, and plot them in properly rescaled length and time units by using the values of the parameters $d_T$ and $\tau_e$ reported in Sec.~3.4.
Fig.~\ref{fig:g1g3} demonstrates the correctness of the rescaling procedure and, in line with previous works~\cite{halverson2011molecular-dynamics,ElhamPRE2021}, the good agreement between our simulations (symbols) and the lattice-tree predictions (dashed black lines), in particular for melts of rings with $\kappa_{\rm bend}=2$ and $Z\approx 50$.
Notice though that our data match well also the predictions of the FLG model~\cite{PanyukovRubinsteinMacromolecules2016} by Ge {\it et al.} (dashed red line in the top-right panel), since the two scaling exponents ($0.29$ {\it vs.} $0.26$, respectively) are within $10\%$ of each other and, hence, beyond the present accuracy of our data.
We investigate now the implications of threading and double-folding for ring dynamics.

The formation of topological links via threadings has been hold responsible of the unique rheological properties of melts of rings as, for instance, the unusually strong extension-rate thickening of the viscosity in extensional flows~\cite{OConnorPRL2020}.
Notably, it was conjectured~\cite{Turner2013} that inter-ring threadings should form a network of topological obstacles which, by percolating through the entire melt, should sensibly slow down the relaxation of the system, not dissimilar from what happens in those materials undergoing the glass transition.
While direct proof of this {\it topological glass} is currently missing, recently Michieletto and others~\cite{Michieletto2016,MichielettoNahaliRosa2017} gave indirect evidence of this through the following numerical experiment: they froze (pinned) a certain fraction $f_p$ of rings in the melt and reported that the dynamics of the remaining ones is considerably reduced if not frozen at all.
In corresponding melts of linear chains this effect is not seen, so they attributed the observed slowing down to the presence of threadings present in rings but absent in linear chains.

\begin{figure*}
\includegraphics[width=1.0\textwidth]{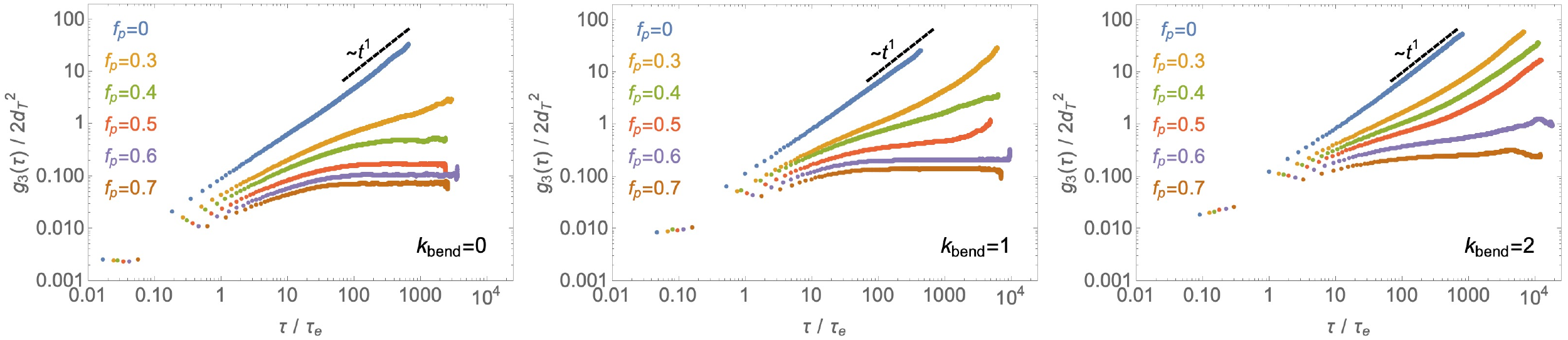}
\caption{\label{fig:g3TopoGlass}
Effects of random pinning on chain dynamics. Chain center of mass time mean-square displacement, $g_3(\tau)$ (Eq.~\eqref{eq:Introduce-g3}), at different pinning fractions, $f_p$, and chain flexibilities, $\kappa_{\rm bend}$ (see legends).
Results correspond to melts with the same number of entanglements, $Z\approx 8$, per individual chain.
}
\end{figure*}
\begin{figure*}
\includegraphics[width=1.0\textwidth]{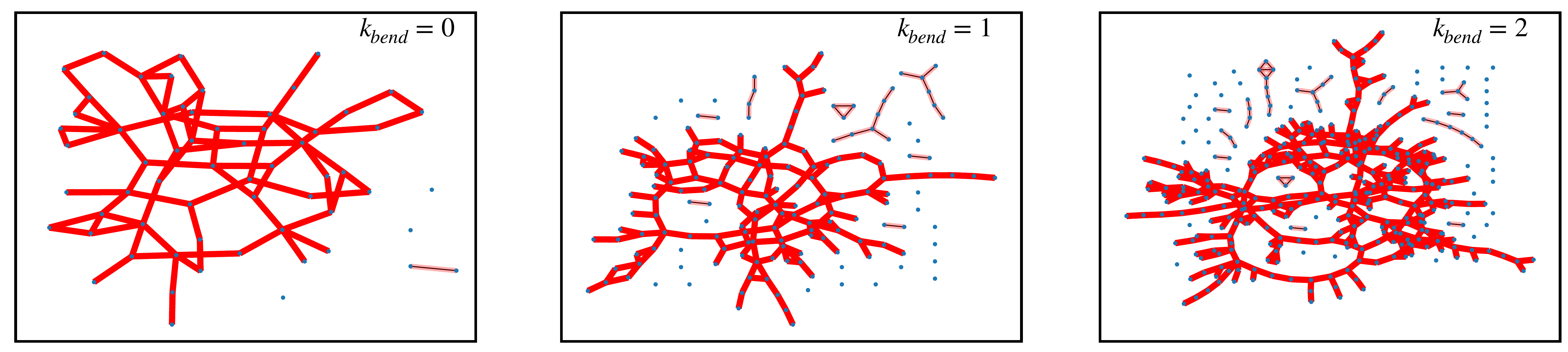}
\caption{\label{fig:RingsNetwork}
Network representation of ring melts with the same number of entanglements, $Z\approx 8$, per individual chain and for different chain flexibilities $\kappa_{\rm bend}$ (see legends).
Each node of the network represents a given ring of the melt and the bond between two nodes means that one of the two ring is threading the (minimal surface of the) other (see the discussion in Sec.~4.1).
For constructing the networks, we have included threading lengths $L_t > L_e$ for which universal scaling behavior emerges (see Sec.~4.1).
}
\end{figure*}

In the numerical experiments discussed here, while double-folding is ``enhanced'' (Sec.~4.1) by the chain local stiffness, long threadings, with $L_t > L_e$ , which should be the ones affecting the dynamics of the system, result to be universal.
At this level, it is unclear what is the impact of these features on the dynamics and whether or not they can be distinguished.
To get some insight into this question, we have taken inspiration from the mentioned pinning numerical experiments~\cite{Michieletto2016,MichielettoNahaliRosa2017} and performed additional simulations of melts of rings for $\kappa_{\rm bend}=0, 1, 2$ with the same number of entanglements $Z=8$ (to ensure the same large-scale behavior, see Fig.~\ref{fig:RgUntangledRings}) and for ring pinning fractions $f_p$ from $0$ ({\it i.e.} no pinning) to $70$\%.
Then we monitored the mean-square displacement, $g_3(\tau)$, of the center of mass of the non-pinned rings.
As shown in Fig.~\ref{fig:g3TopoGlass}, our results are consistent with the proposed picture: more flexible rings ($\kappa_{\rm bend}=0$, left panel) which we have interpreted as less double-folded are already completely frozen at $f_p\simeq30-40$\% while the stiffer (and comparably more double-folded) rings ($\kappa_{\rm bend}=2$, right panel) require $f_p\simeq60-70$\%.
Importantly, these specific fractions do not seem to depend on the finite size of our systems: dynamic runs with smaller (and, arguably, more finite-size dependent) systems give similar values for the freezing fractions (see Fig.~S7 in SI). 
Finally, we see that this is compatible with a network of percolating topological constraints.
To this purpose, we represent the melts as networks where each node corresponds to a ring and we draw a link between two nodes ({\it i.e.}, rings) whenever the minimal surface of one of the two rings is pierced by a threading segment of the other ring of contour length $L_t>L_e$ ({\it i.e.}, we neglect the shallow threadings).
A picture of these networks is shown in Fig.~\ref{fig:RingsNetwork}: it is seen that, at increasing chain stiffness, the fraction of rings globally connected through threadings sensibly decreases with chain stiffness, from\footnote{In principle, these percentages may be affected by finite-size effects due to the limited number of chains of our melts. Nonetheless, our conclusions should be fairly robust.} $\sim 93\%$ for $\kappa_{\rm bend}=0$ to $\sim74\%$ for $\kappa_{\rm bend}=1$ and $\kappa_{\rm bend}=2$, {\it i.e.} the network of threadings is percolating less through the stiffer melts.
Notice that this decrease corresponds roughly to the fractional amount ($\approx 20\%$) of rings which need to be additionally pinned in the stiffer systems compared to $\kappa_{\rm bend}=0$ in order to get complete freezing (Fig.~\ref{fig:g3TopoGlass}).
To summarize, flexible melts showing higher propensity to shallow threadings (Fig.~\ref{fig:ntvsZ}) are more strongly affected by the pinning, pointing to a possible dynamical role of the shallow threadings.
In contrast -- yet consistent with the pinning fraction -- the deep threadings in flexible melts form more connected networks in comparison to the stiffer melts seemingly suggesting that these alone are the relevant ones in the pinning experiment.
The latter result is however surprising, because the deep threadings were found to behave universally for the systems with different stiffness (see Fig.~\ref{fig:ntvsZ}).
We leave this puzzle for future work.

Another view on the results of our pinning experiments shows that the threading constraints are a relevant physical feature that needs to be taken into account to refine models for the ring dynamics.
As we mentioned~\eqref{eq:tau_relax-FLG}, the ring relaxation time with ($\theta=1$) or without ($\theta=0$) the tube dilation can be written~\cite{PanyukovRubinsteinMacromolecules2016} as
$$
\tau_r \sim \tau_e (N/Ne)^{2+ (1-\theta)\nu/\nu_{\rm path} + \theta\nu} \, .
$$
Notably, the pinning of a large fraction $f_p$ of rings must stop the dilation, because the pinned chains around a mobile ring cannot move, hence impose constraints at all times. The plateau of the $g_3$ that we observe indicates a divergence of the relaxation time, but the FLG as well as the lattice models give a finite prediction for the relaxation time even for inhibited dilation $\theta=0$. Indeed this is the consequence of neglecting of the threadings. Although the FLG model is built on the full tube dilation in the melt, and therefore, its comparison to the pinning with a limited dilation is problematic, its general formulation~\eqref{eq:tau_relax-FLG} allows to consider a partial ($\theta<1$) or no tube dilation.
Last but not the least, the fact that both, the lattice-tree and the FLG model, underestimate the scaling exponent of the diffusion coefficient with $N$ in comparison to simulations and experiment: a similar underestimation of the scaling exponent of the diffusion coefficient is found in \emph{linear} melts when compared to a na\"ive reptation theory~\cite{DoiEdwardsBook}.
The agreement between the theory and the experiment is restored when contour length fluctuations and tube dilation are incorporated.
In the rings, the FLG model already includes the tube dilation, yet the prediction of the exponent does not agree with the experiment, which indicates that this is due to the neglect of the threadings.
Another explanation of the discrepancy might be the correlation hole effect as indicated in~\cite{KrutevaPRL2020}. Whether this is independent of threadings remains to be elucidated.

\section{5. Discussion and conclusions\label{sec:DiscConcls}}
The conformational properties of unknotted and non-concatenated ring polymers in dense melts represent one of the remaining unsolved challenge in polymer physics.
In this work, in particular, we have focused on which properties of the rings can be interpreted as signatures of the hypothesis proposed long ago~\cite{KhokhlovNechaev1985,Rubinstein1986,Rubinstein1994} that the polymer fiber double-folds around a branched (lattice-tree) path.

In this respect, global observables like the polymer mean-square gyration radius or the mean-square magnetic radius ($\langle R_g^2\rangle$ and $\langle R_m^2\rangle$, Fig.~\ref{fig:RgUntangledRings}) are very useful for model validation but otherwise offer little insight.

On the contrary, robust evidence for double-folding comes by exploiting the mean polymer shape (Fig.~\ref{fig:LambdaSq1vsLambdaSq3etc}).
The $3$ principal axes of the polymer are very different in size and, for relatively moderate polymer lengths, they are not at the same ``point'' of the crossover to asymptotic behavior:
in particular, on the studied range where the largest and the smallest axes are not proportional to each other, we have shown that their functional relation is in {\it quantitative} agreement with the lattice-tree model.
This argument is proposed here for the first time, and it could be useful to revise data relative to other polymer models at its light.

Additional signatures of double-folding on polymer contour lengths $z = \ell / L_e \lesssim 1$ become manifested also in the characteristic negative well of the bond-vector orientation correlation function (Fig.~\ref{fig:tgtgCorrel}) and in the ``softer'' slope of the mean contact probability $\langle p_c(\ell)\rangle$ (Fig.~\ref{fig:pc}) displayed before the asymptotic $\sim \ell^{-1}$ power-law behavior takes effectively place. 
Intriguingly, the reported $\langle p_c(\ell)\rangle$ for chromatin fibers measured in conformation capture experiments~\cite{Dekker_HiC2009} displays the same systematic two-slope crossover~\cite{Fudenberg2017} for $\ell \lesssim 10^5$ basepairs, {\it i.e.} below the estimated~\cite{RosaEveraersPlos2008} entanglement length of the chromatin fiber. 
Notice that, in this respect, the formulated hypothesis that such ``shoulder'' in the contact probability derives from active loop-extrusion~\cite{Fudenberg2017} provides a dynamic explanation about how (double-folded) loops can form, otherwise it is not needed {\it per se} in order to explain the observed trend in the contact probabilities. 

Overall it is worth stressing that these features, that we interpret as manifestations of double-folding, can be made explicit only by introducing some (even if only moderate) bending penalty to the polymer elasticity.

Using minimal surfaces we confirm (Fig.~\ref{fig:minSvsN}) 
that rings may reduce their threadable surfaces via double-folding on the entanglement scale and that the piercings of the minimal surface occur only on the scale below the entanglement tube radius.
Yet, the threading loops can be numerous and their length can exceed the entanglement length, but these threading features (Fig.~\ref{fig:ntvsZ} and Fig.~S3) evaluated on length scales larger than one entanglement length behave universally.
The universal behavior shows the applicability and the relevance~\cite{Ge_crazing_2021} of the (linear) entanglement scale to ring polymer melts.

From the point of view of ring dynamics, we have reported that single monomer and global chain motions are in good agreement (Fig.~\ref{fig:g1g3}) with the theory stating that the dominant mode of relaxation is mass transport along the mean path of the underlying tree.
Yet the threading constraints can cause divergence of the relaxation time if a fraction of rings is immobilized -- a feature not predicted by the theories. Revisiting the theories to incorporate threading constraints explicitly might also help to clarify if the conjecture on the existence of topological glass in equilibrium is valid.

\section{Supporting Information}
Table of mean values of structural properties of ring polymers, monomer mean-square displacement $g_2$ in the frame of the ring center of mass, ratios of mean-square gyration radius and mean-square magnetic radius, mean minimal surface after removing the contribution of shallow threadings, distribution functions of distances $d$ between subsequent piercings, mean contact probabilities, mean-square eigenvalues of the gyration tensor, effect of random pinning on smaller melts.

%
\begin{acknowledgement}
JS and AR acknowledge networking support by the COST Action CA17139 (EUTOPIA).
The computational results presented have been achieved in part using the Vienna Scientific Cluster (VSC).
\end{acknowledgement}

\providecommand{\latin}[1]{#1}
\makeatletter
\providecommand{\doi}
  {\begingroup\let\do\@makeother\dospecials
  \catcode`\{=1 \catcode`\}=2 \doi@aux}
\providecommand{\doi@aux}[1]{\endgroup\texttt{#1}}
\makeatother
\providecommand*\mcitethebibliography{\thebibliography}
\csname @ifundefined\endcsname{endmcitethebibliography}
  {\let\endmcitethebibliography\endthebibliography}{}

\clearpage

\begin{center}
\textbf{\Large Entanglement length scale separates threading from branching of unknotted and non-concatenated ring polymers in melts \\ \vspace*{1.5mm} -- Supporting Information --} \\
\vspace*{5mm}
Mattia Alberto Ubertini, Jan Smrek, Angelo Rosa
\vspace*{10mm}
\end{center}

\setcounter{equation}{0}
\setcounter{figure}{0}
\setcounter{table}{0}
\setcounter{page}{1}
\setcounter{section}{0}
\makeatletter
\renewcommand{\theequation}{S\arabic{equation}}
\renewcommand{\thefigure}{S\arabic{figure}}
\renewcommand{\thetable}{S\arabic{table}}
\renewcommand{\thesection}{S\arabic{section}}

\clearpage

\makeatletter
\@fpsep\textheight
\makeatother

\begin{table*}
\begin{adjustbox}{width=1\textwidth}
\small
\begin{tabular}{cccccccc}
$N$ & $M$ & $\langle R_g^2 \rangle$ & $\langle R_m^2 \rangle$ & $\langle {\rm minS} \rangle$ & $\langle \Lambda_1^2 \rangle$ & $\langle \Lambda_2^2 \rangle$ & $\langle \Lambda_3^2 \rangle$ \\
\hline
\hline
\\
\\
\multicolumn{8}{c}{$\kappa_{\rm bend}=0$} \\
\hline
\\
$40$ & $1000$ & $3.3007 \pm 0.0007$ & $1.8937 \pm 0.0008$ & $11.60 \pm 0.05$ & $2.1160 \pm 0.0007$ & $0.8280 \pm 0.0003$ & $0.35669 \pm 0.00008$ \\
$80$ & $500$ & $6.304 \pm 0.004$ & $3.447 \pm 0.004$ & $25.6 \pm 0.2$ & $4.052 \pm 0.004$ & $1.560 \pm 0.001$ & $0.6912 \pm 0.0003$ \\
$160$ & $250$ & $11.76 \pm 0.02$ & $6.19 \pm 0.02$ & $55.6 \pm 0.4$ & $7.56 \pm 0.02$ & $2.891 \pm 0.005$ & $1.310 \pm 0.001$ \\
$320$ & $125$ & $21.48 \pm 0.08$ & $11.0 \pm 0.1$ & $119.7 \pm 1.2$ & $13.74 \pm 0.07$ & $5.29 \pm 0.02$ & $2.447 \pm 0.005$ \\
$640$ & $62$ & $38.6 \pm 0.2$ & $18.9 \pm 0.2$ & $256.2 \pm 2.8$ & $24.7 \pm 0.2$ & $9.37 \pm 0.05$ & $4.47 \pm 0.01$ \\
\hline
\\
\\
\multicolumn{8}{c}{$\kappa_{\rm bend}=1$} \\
\hline
\\
$40$ & $1000$ & $4.133 \pm 0.001$ & $2.262 \pm 0.002$ & $12.82 \pm 0.07$ & $2.772 \pm 0.001$ & $0.9778 \pm 0.0004$ & $0.38363 \pm 0.00008$ \\
$80$ & $500$ & $7.915 \pm 0.008$ & $4.09 \pm 0.01$ & $28.6 \pm 0.2$ & $5.220 \pm 0.006$ & $1.893 \pm 0.002$ & $0.8028 \pm 0.0004$ \\
$160$ & $250$ & $14.52 \pm 0.02$ & $7.19 \pm 0.05$ & $61.9 \pm 0.6$ & $9.43 \pm 0.02$ & $3.506 \pm 0.007$ & $1.573 \pm 0.002$ \\
$234$ & $171$ & $19.97 \pm 0.06$ & $9.70 \pm 0.04$ & $94.3 \pm 0.8$ & $12.95 \pm 0.03$ & $4.834 \pm 0.008$ & $2.223 \pm 0.002$ \\
$320$ & $125$ & $25.8 \pm 0.1$ & $12.38 \pm 0.07$ & $132.3 \pm 1.4$ & $16.67 \pm 0.08$ & $6.28 \pm 0.02$ & $2.936 \pm 0.006$ \\
$640$ & $62$ & $44.3 \pm 0.5$ & $20.5 \pm 0.3$ & $279.7 \pm 3.3$ & $28.1 \pm 0.4$ & $10.73 \pm 0.09$ & $5.20 \pm 0.03$ \\
\hline
\\
\\
\multicolumn{8}{c}{$\kappa_{\rm bend}=2$} \\
\hline
\\
$40$ & $1000$ & $5.600 \pm 0.006$ & $2.935 \pm 0.005$ & $14.8 \pm 0.1$ & $3.998 \pm 0.004$ & $1.199 \pm 0.002$ & $0.4030 \pm 0.0003$ \\
$80$ & $500$ & $10.82 \pm 0.02$ & $5.15 \pm 0.03$ & $33.0 \pm 0.3$ & $7.45 \pm 0.02$ & $2.426 \pm 0.007$ & $0.946 \pm 0.002$ \\
$104$ & $385$ & $13.56 \pm 0.02$ & $6.27 \pm 0.02$ & $44.1 \pm 0.4$ & $9.21 \pm 0.01$ & $3.092 \pm 0.004$ & $1.261\pm 0.001$ \\
$160$ & $250$ & $19.4 \pm 0.2$ & $8.66 \pm 0.06$ & $70.7 \pm 0.7$ & $12.87 \pm 0.07$ & $4.50 \pm 0.02$ & $1.943 \pm 0.006$ \\
$320$ & $125$ & $33.2 \pm 0.3$ & $14.2 \pm 0.1$ & $148.7 \pm 1.8$ & $22.0 \pm 0.3$ & $7.90 \pm 0.07$ & $3.65 \pm 0.02$ \\
$640$ & $62$ & $54.9 \pm 1.4$ & $22.4 \pm 0.4$ & $308.8 \pm 3.5$ & $35.6 \pm 1.3$ & $12.9 \pm 0.2$ & $6.36 \pm 0.06$ \\
\hline
\hline
\end{tabular}
\end{adjustbox}
\caption{
\label{tab:avRg2-etc}
Conformational properties of melts of ring polymers with bending stiffness $\kappa_{\rm bend}$.
$N$: number of monomers per chain;
$M$: number of chains in the melt;
$\langle R_g^2 \rangle$: mean-square gyration radius (Eq.~(19) in the main paper); 
$\langle R_m^2 \rangle$: mean-square magnetic radius (Eq.~(20) in the main paper); 
$\langle {\rm minS} \rangle$: mean minimal surface; 
$\langle\Lambda_\alpha^2\rangle$: mean value of the $\alpha$-th eigenvalue ($\alpha=1,2,3$) of the gyration tensor ${\mathcal G}_{\alpha\beta}$ (Eq.~(26) in the main paper); 
The reported values with the corresponding error bars have been rounded to the first decimal digit.
The data for $\langle {\rm minS} \rangle$ are for a coarse resolution of the triangulation procedure adopted to calculate the minimal surface of a ring polymer (see Ref.~\citep{SmrekGrosbergACSMacroLett2016} for the technical details).
A finer triangulation gives essentially the same values (``$\circ$-symbols'' vs. ``$\times$-symbols'' in Fig.~5 in the main paper) 
which we have not reported explicitly in this Table.
}
\end{table*}

%
\begin{figure*}
$$
\begin{array}{ccc}
\includegraphics[width=0.33\textwidth]{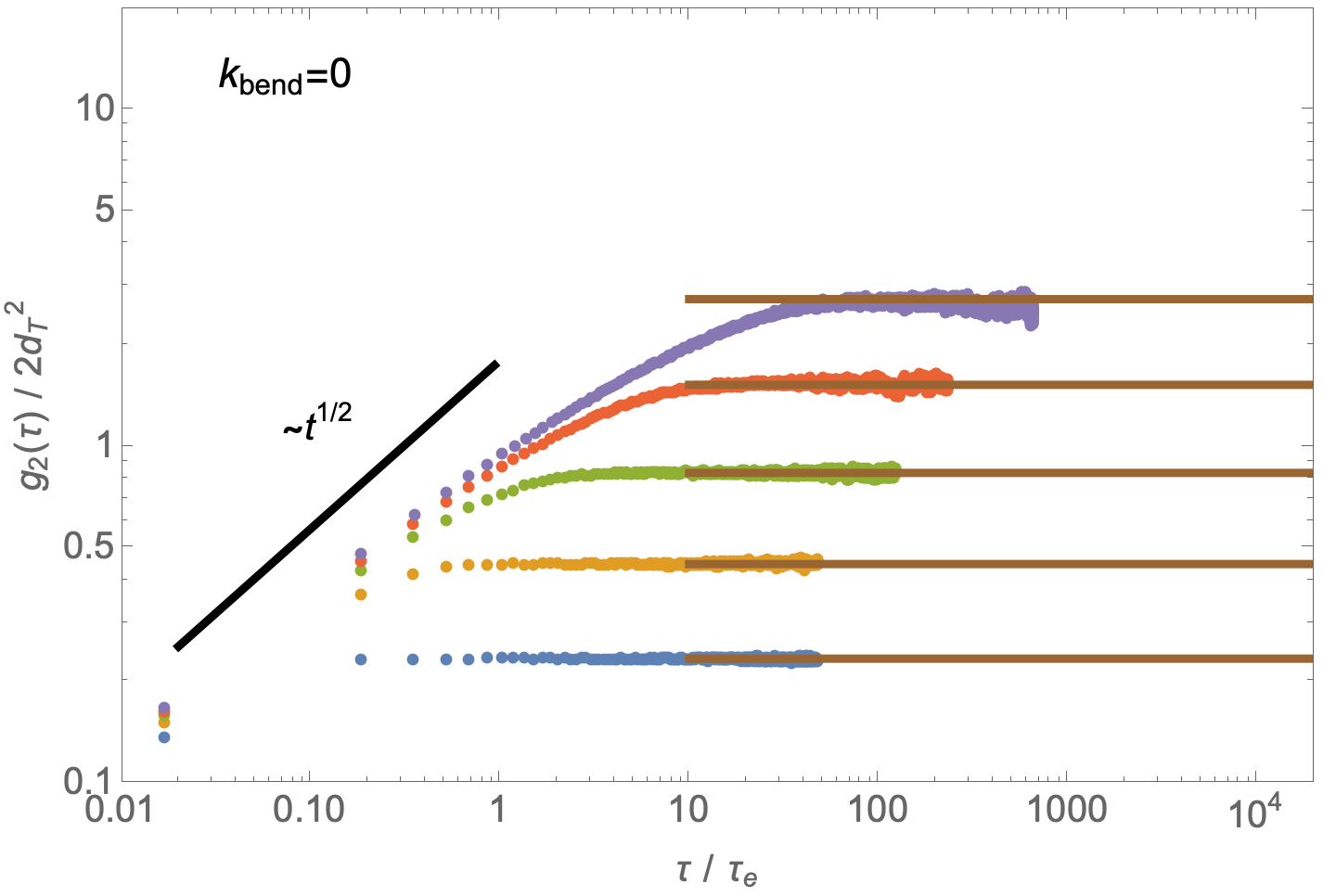} & \includegraphics[width=0.33\textwidth]{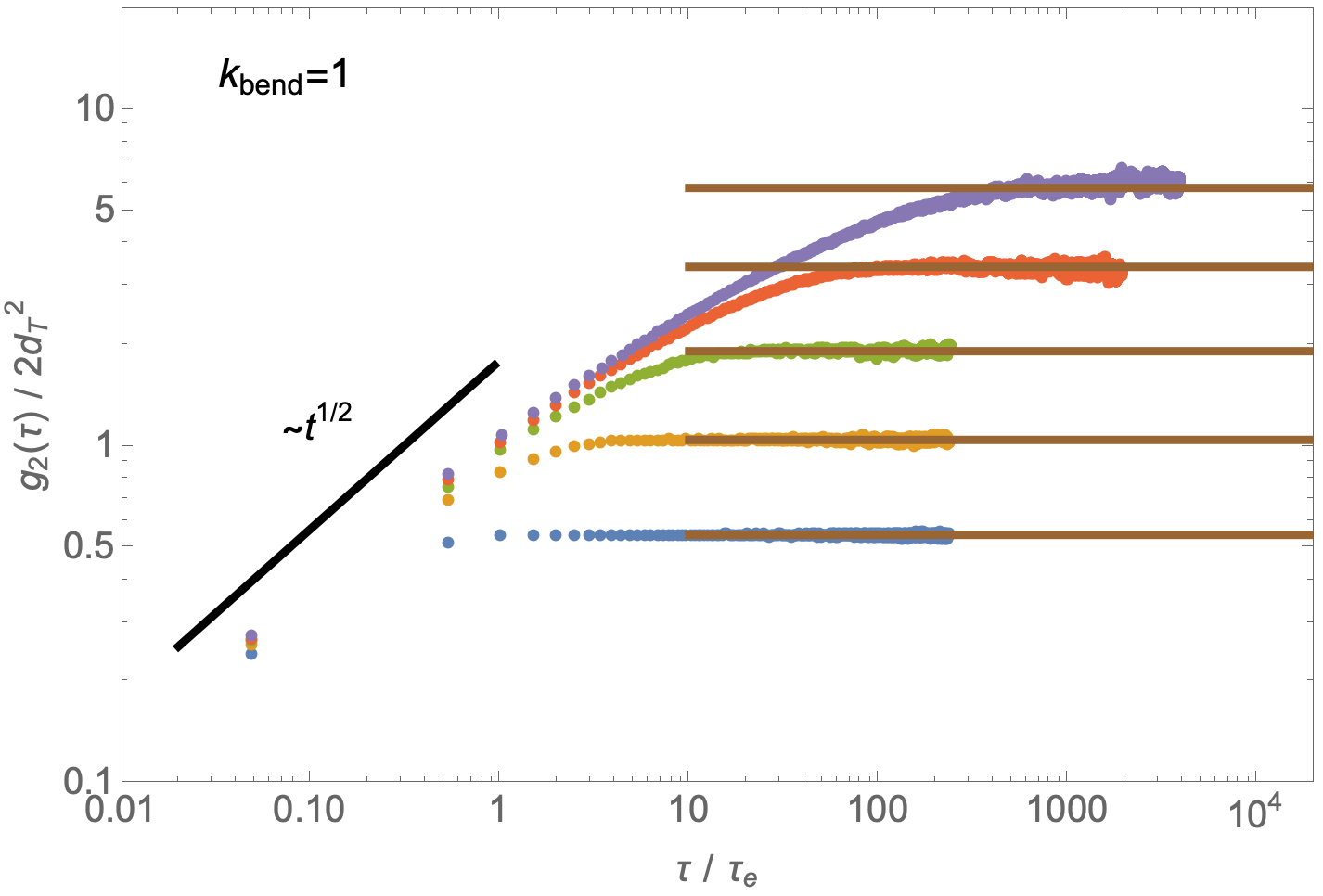} & \includegraphics[width=0.33\textwidth]{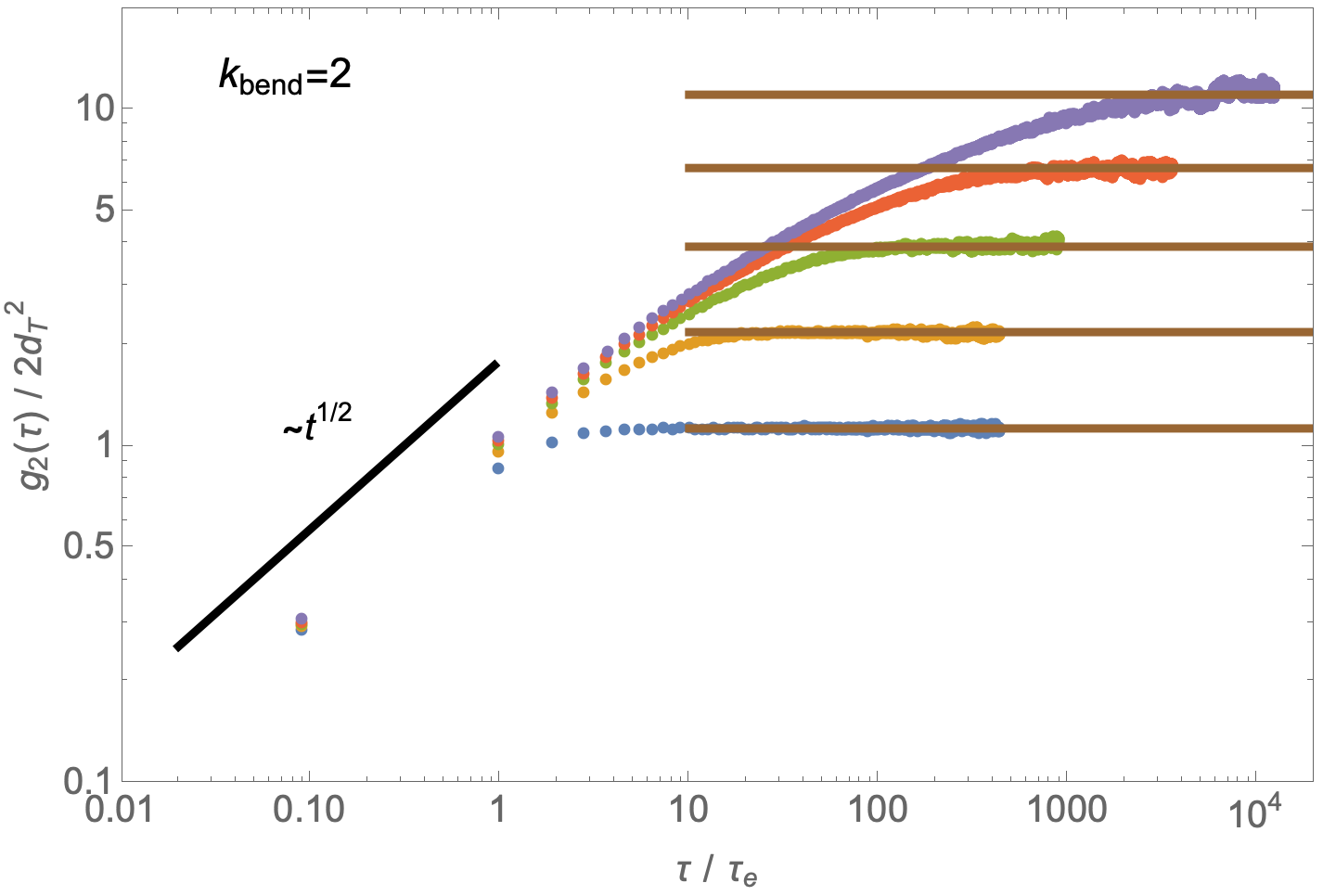}
\end{array}
$$
\caption{\label{fig:g2}
Monomer time mean-square displacement, $g_2(\tau)$ (see Eq.~(12) in the main paper), 
in the frame of the centre of mass of the corresponding chain.
The horizontal lines are for the stationary values, $=2\langle R_g^2 (N) \rangle$ (see Sec.~2.3 in the main paper), 
reported in Table~\ref{tab:avRg2-etc}.
The dashed black lines correspond to the initial Rouse behavior where $g_2(\tau) \sim g_1(\tau)$ (see Eq.~(10) in the main paper). 
Color code is as in Fig.~10 in the main paper. 
}
\end{figure*}
\begin{figure*}
\includegraphics[width=0.48\textwidth]{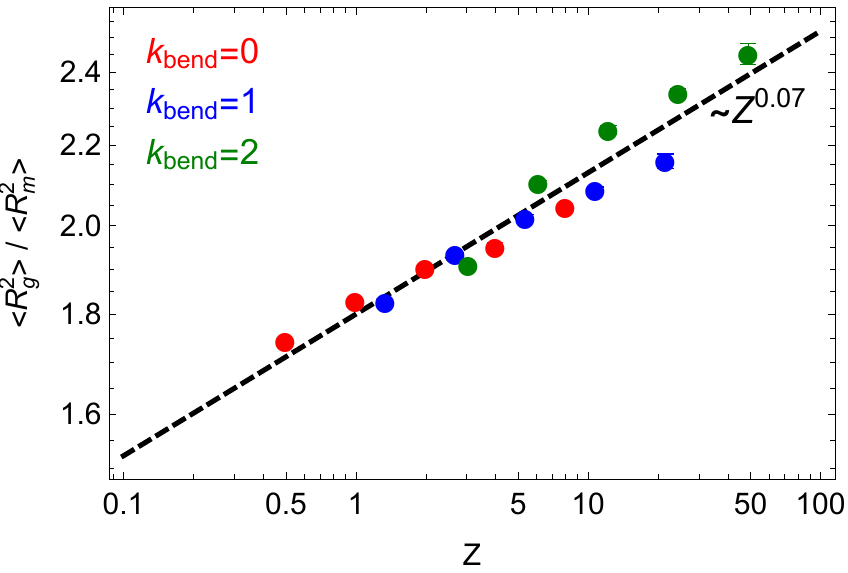}
\caption{\label{fig:Rg2vsRm2}
$\langle R_g^2 \rangle / \langle R_m^2 \rangle$: ratio of the mean-square gyration radius (Eq.~(19) in the main paper) 
to the mean-square magnetic radius (Eq.~(20) in the main paper) 
as a function of the total number of entanglements, $Z=L/L_e$, of the chain.
Notice the {\it weak} increasing of the ratio with $Z$ (the straight line is obtained by best fit of the data to a simple power-law).
}
\end{figure*}
\begin{figure*}
\includegraphics[width=0.45\textwidth]{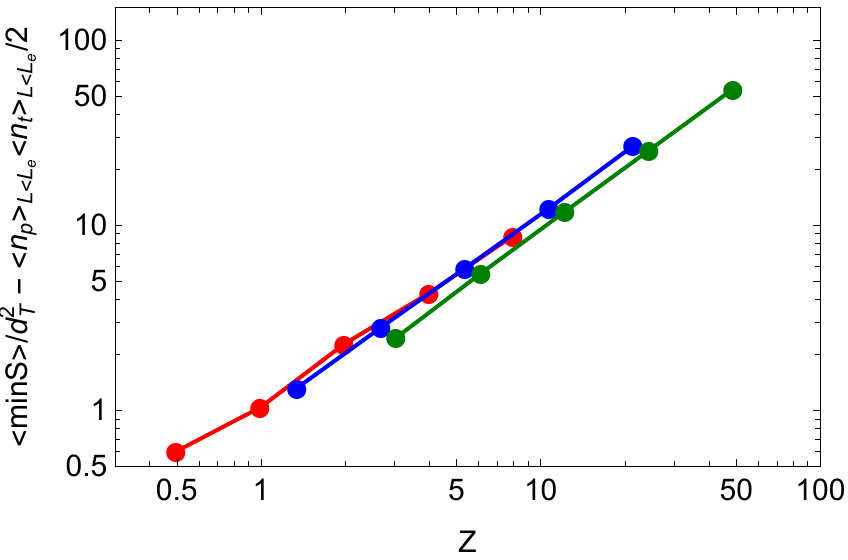}
\caption{\label{fig:minS-minus-ShallowThreadings}
Mean minimal surface area, $\langle {\rm minS}\rangle$, of ring polymers as a function of the total number of entanglements, $Z$, of the chain after subtracting the ``contribution'' to the area by short (or, {\it shallow}) threadings with contour length $L<L_e$.
This contribution is calculated by assuming that each threading span an area $\sim d_T^2$
and there are $\sim \langle n_p \rangle_{L<L_e} \langle n_t\rangle_{L<L_e} /2$ of such threadings spanning the minimal surface of each ring.
$\langle n_t\rangle_{L<L_e}$ and $\langle n_p\rangle_{L<L_e}$ are (as in Fig.~6 in the main text, but here restricted to shallow threadings) the mean number of rings threaded by a single ring (which equals the mean number of threadings received by a single ring) and the mean number of times a ring penetrates the minimal surface of any other ring, while the ``$1/2$''-factor takes into account the fact that a single threading loops counts as twice in terms of contacts on the minimal surface.
}
\end{figure*}
\begin{figure*}
\includegraphics[width=0.5\textwidth]{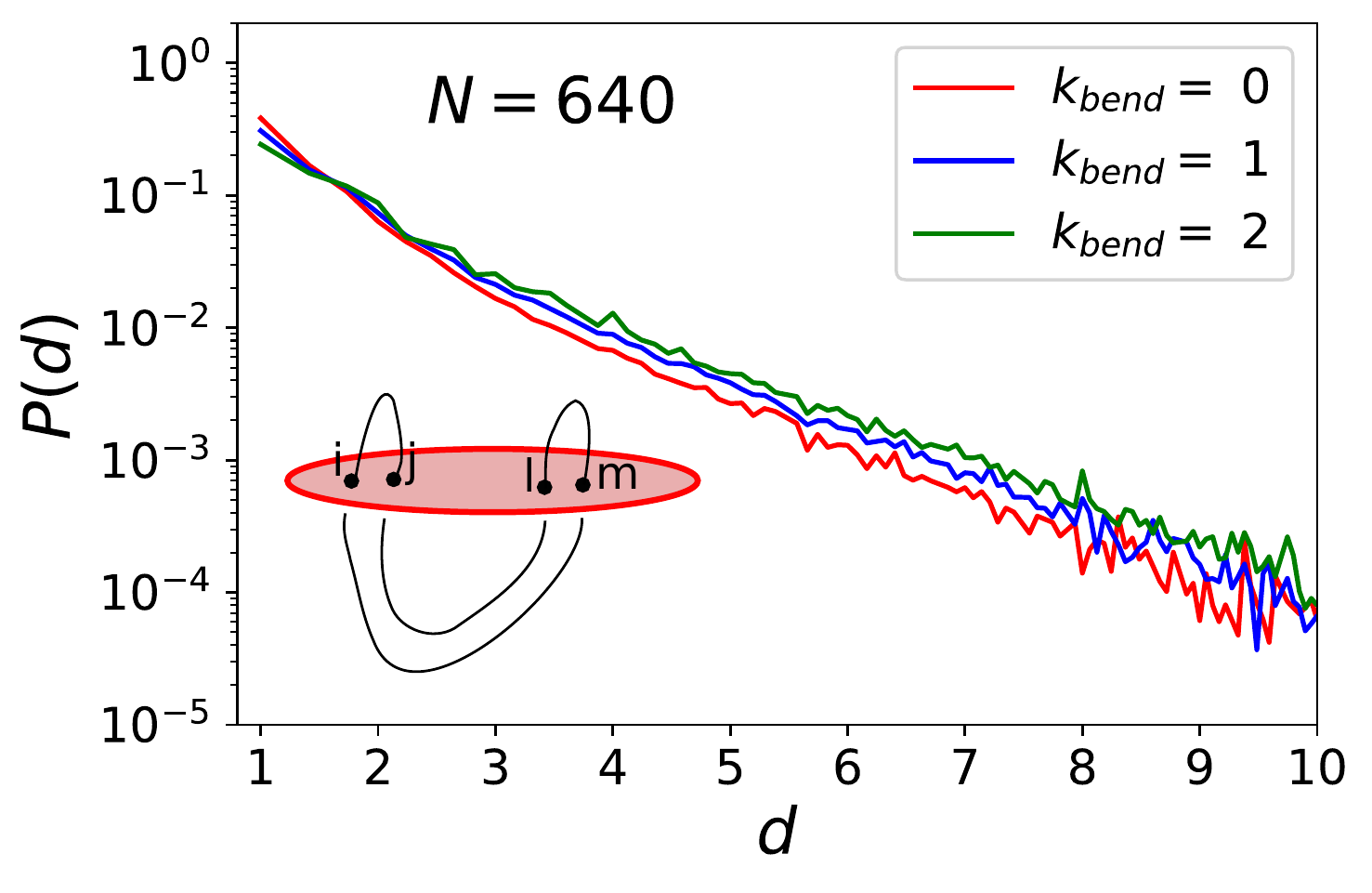}
\caption{\label{fig:pd_distro}
Distribution functions of distances $d$ between subsequent piercings on the minimal surfaces of rings with $N=640$ and different flexibilities.
Clearly the distributions are exponential. The mean values are $\langle d \rangle =  1.87$, $2.1$ and $2.33$ (in units of lattice site $a$) for $k_{\rm bend}=0$, $k_{\rm bend}=1$ and $k_{\rm bend}=2$ respectively.
Interestingly, in contrast to the tube diameter $d_{T}$, $\langle d \rangle$ grows with the stiffness. Note that, in the case of more than two piercings per surface, there are two different sets of such distances depending on the parity of the piercings (see the inset).
For example (see the sketch in the inset) if four bonds $i$,$j$,$l$,$m$ are piercing, one can consider the distances $d(i,j)$ ({\it i.e.}, the distance between bonds $i$ and $j$) and $d(l,m)$, or alternatively, $d(j,l)$ and $d(i,m)$.
In such cases, we take into account the set with smaller mean ($d(i,j)$ and $d(l,m)$ for the case in the inset). The reason behind this choice is that we want to use $\langle d \rangle^{2}/2$ as a proxy for the area per piercing.
Hence if pairs of piercings are distant from each other (as is the case of $(i,j)$ being far from $(l,m)$), they represent two uncorrelated events ($(i,j)$ being uncorrelated to $(l,m)$) and the square of the distance between them ({\it e.g.}, $d(j,m)$) does not represent the increase of the minimal surface.
}
\end{figure*}
\begin{figure*}
\includegraphics[width=0.45\textwidth]{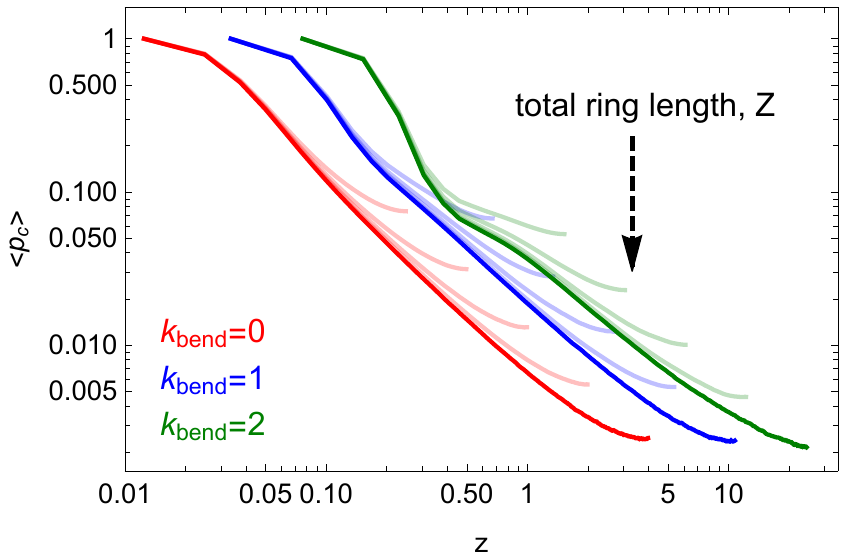}
\caption{\label{fig:pc}
Mean contact probabilities, $\langle p_c(\ell) \rangle$, as a function of the contour length separation in number of entanglements, $z\equiv \ell/L_e$. 
Lines in full colors are for the longest rings ($N=640$) while lines in fainter colors are for chains of shorter contour lengths (see arrow's direction).
}
\end{figure*}
\begin{figure*}
\includegraphics[width=0.48\textwidth]{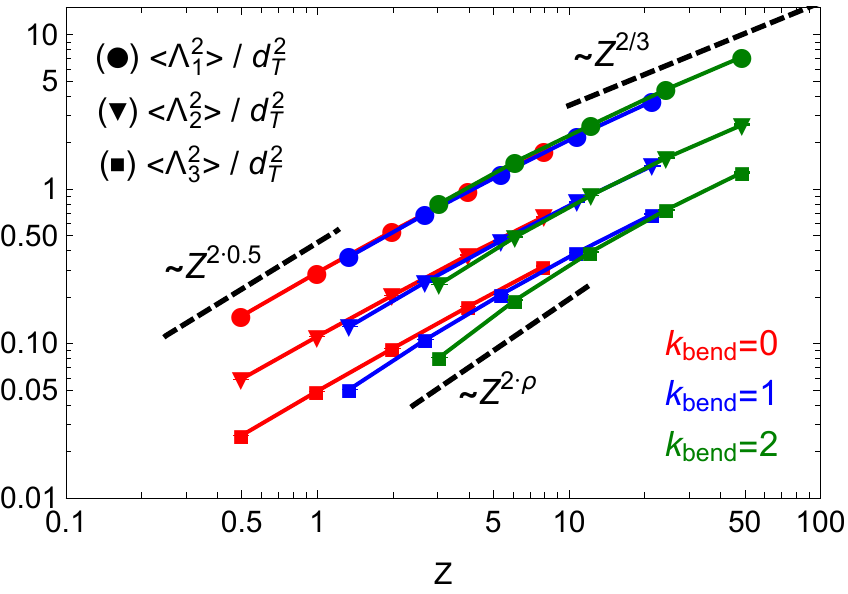}
\caption{\label{fig:AsymmEigenvals}
Scaling behavior of the mean-square eigenvalues $\langle \Lambda_{\alpha=1,2,3}^2 \rangle$ of the gyration tensor (Eq.~(26) in the main paper) 
of untangled ring polymers as a function of the total number of entanglements, $Z=L/L_e$, of the chain.
Notice the large finite-size corrections to scaling for the smallest eigenvalue $\langle \Lambda_3^2 \rangle$ and its behavior $\sim Z^{2\rho} \sim \langle L_{\rm tree}\rangle^2$ (Eqs.~(6) and~(28) in the main paper) 
for small to moderate number of entanglements per chain $Z$.
Error bars are smaller than the symbols size.
}
\end{figure*}
%

\begin{mcitethebibliography}{66}
\bibitem[Mot()]{MotivationForLk}
From Eqs.~\eqref{eq:LoverLe=Z} and~\eqref{eq:elloverLe=z} we see a further
  reason of interest to study melts of rings by varying the Kuhn length
  $\ell_K$: if the prefactor was stiffness-dependent, the universal scaling of
  all the quantities in terms of the number of entanglements would not be
  expected.\relax
\mciteBstWouldAddEndPunctfalse
\mciteSetBstMidEndSepPunct{\mcitedefaultmidpunct}
{}{\mcitedefaultseppunct}\relax
\EndOfBibitem
\end{mcitethebibliography}
\end{document}